\documentclass[11pt,nofootinbib,aps,preprint]{revtex4-1}

\usepackage{amsmath,mathrsfs}
\usepackage{graphicx,epsfig}
\usepackage{latexsym,amssymb}
\usepackage{epsf}
\usepackage{ifpdf,natbib,lineno}
\usepackage{xcolor}
\usepackage{multirow}

\def\sech{\hspace{0.75mm}{\rm sech}}

\begin{document}

\title{{\bf Introducing a  Relativistic Nonlinear Field System With a Single Stable Non-Topological  Soliton Solution in $1+1$ Dimensions}}

\author{ M. Mohammadi$^{*}$} \address{Physics Department, Persian Gulf University, Bushehr, 75169, Iran.}
\date{\today}
\email{physmohammadi@pgu.ac.ir}

\begin{abstract}

In this paper we present a new  extended  complex nonlinear  Klein-Gordon Lagrangian density,   which bears a single   non-topological    soliton  solution    with a specific   rest frequency $\omega_{s}$ in $1+1$ dimensions. There is a proper  term in the new Lagrangian density, which behaves like a massless spook  that surrounds  the  single soliton solution  and  opposes  any internal  changes. In other words,  any arbitrary variation in the single soliton   solution leads to an increase in the total energy.
Moreover, just for the single  soliton solution, the general  dynamical equations are reduced to  those versions of a special type of the standard well-known complex nonlinear Klein-Gordon systems, as its dominant dynamical equations.

\end{abstract}

\maketitle


\section{Introduction}

Study of soliton solutions in  relativistic classical field theories is an attempt to model particles in terms of non-singular, localized solutions of properly tailored nonlinear PDEs \cite{rajarama,Das,lamb,Drazin,TS}. Kink  and anti-kink solutions of the real nonlinear Klein-Gordon (KG) equations in $1+1$ dimensions were a successful effort  to this end  \cite{phi41,phi42,phi43,phi44,phi45,OV,GH,MM1,MR,JRM1,JRM2,DSG1,DSG2,DSG3,MM2,waz,ana1,
ana2,Kink1,Kink2,Kink3,Kink4,Kink5,Kink6,Kink7,Kink8,Kink9,Kink10,Kink11,Kink12,Kink13,
Kink14,Kink15,Kink16,Kink17}. In this context, the recent interesting and important results on kink-(anti)kink interactions in models which possess kinks with power-law tails (power-law asymptotical behavior
of the kink solutions), can be mentioned \cite{Kink8,Kink9,Kink10,Kink11,Kink12}. Moreover, in the context of kinks and their interactions it is also worth to mention recent results on the $\phi^8$ and more complex models \cite{Kink13,Kink14,Kink15,Kink16,Kink17}.
Solitons are, in some respects, similar to physical particles. They  satisfy  the relativistic  energy-rest mass-momentum relation and are  stable objects. Stability is the main condition for a solitary wave solution to be a soliton\footnote{According to some well-known references such as  \cite{rajarama}, the  stability is just  a necessary condition for a solitary wave solution to be a soliton; more precisely,  a solitary wave solution is a soliton if it reappears  without any distortion after collisions. In this paper,  we only accept the stability condition for the definition of a soliton solution.}. As regards stability, there are many different criteria. The first and foremost criterion is to examine whether a solitary wave solution is topological or non-topological.   Basically, the topological solitary wave solutions are inevitably stable, among which, one can mention the kink (anti-kink) solutions \cite{phi41,phi42,phi43,phi44,phi45,OV,GH,MM1,MR,JRM1,JRM2,DSG1,DSG2,DSG3,MM2,waz,ana1,ana2,Kink1,Kink2,Kink3,Kink4,
Kink5,Kink6,Kink7,Kink8,Kink9,Kink10,Kink11,Kink12,Kink13,Kink14,Kink15,Kink16,Kink17} and magnetic monopole solitons of 't Hooft Polyakov model \cite{rajarama,TS,toft,MKP,TOF} and   solitons  of the  Skyrme's model  \cite{TS,SKrme,SKrme2,SKrme3,SKrme4}

For the  non-topological solitary wave  solutions, a known  standard stability  criterion (method) is the Vakhitov-Kolokolov (or the classical) criterion,  which involves  obtaining the permissible small   solutions of the  linearized  equations  of motion  above the background of the solitary wave solutions \cite{Vak1,Vak2,Vak777,Vak3,Vak4,Vak5,Vak6}.  In this method,    we first consider any  permissible small perturbation  as a localized oscillatory  function, as an ansatz,   with a specific   frequency $\omega$, and then try   to find the possible    eigenfunctions and eigenfrequencies. If we find an eigenfunction with a   pure imaginary  eigenfrequency $\omega$, or any growing mode,  then the solitary wave solution is unstable.  If this criterion is used  for the topological kink (anti-kink) solutions,   the existence of the non-trivial internal modes would be  possible  for some kink solutions, which causes such  kinks (anti-kink) to display  a permanent vibrational behavior  in a collision process \cite{phi41,phi45,OV,GH,MM1,MR}. Moreover, for the non-topological solitary wave solutions  of the real one-field  nonlinear Klein-Gordon systems, this criterion  leads us to conclude   that there are not any stable solution at all \cite{Vak3}.

For the   complex nonlinear  Klein-Gordon (CNKG) systems, it was shown that there are some non-topological  solutions that are called Q-balls \cite{Vak3,Vak4,Vak5,Vak6,Lee3,Scoleman,R1,R2,R3,R4,R5,R6,R7,Riazi2,MM3}. In fact, they are  some solitary wave  profiles,  which  can be  identified   with their specific rest frequencies $\omega_{o}$.  In general, it was shown that  Q-balls have the minimum rest energy among the other solutions with the same electrical charge \cite{Vak6,Scoleman}. Based on the Vakhitov-Kolokolov stability criterion, the stability conditions  of  the Q-balls were   obtained  in detail \cite{Vak777,Vak3,Vak4,Vak5,Vak6}. Moreover, some researchers  have tried  to examine  the stability  of such non-topological solutions according to the paradigm of the quantum mechanics \cite{Vak6,Lee3}, which leads to the quantum mechanical stability  criterion. It is   based on a comparing   between the properties of the Q-ball (such as charge and rest energy) and the properties of the free scalar particle quanta. A Q-ball which is quantum-mechanically stable can not decay to  a number of  free quanta.


In this paper we use a new criterion (i.e. the energetically stability criterion \cite{Derrick,PH1,PH2}) for the relativistic field systems with the non-topological solitary wave solutions.  We assume that a  non-topological solitary wave solution is stable if any arbitrary deformation in its internal  structure, when it is at rest, leads to an increase in the related total  energy. In other words, we assume that  a stable solitary wave solution has the minimum energy among the other close solutions. According to this  new criterion, we will show that none of the Q-balls are stable objects. It should be noted that,   this new criterion is  different from  the Vakhitov-Kolokolov criterion, but both are classical. Based on the  Vakhitov-Kolokolov method, we examine  the  dynamical equations of motion for the small oscillations above the background of the solitary wave solutions (and   linearized   them to obtain another eigenvalue equation for the permissible small perturbations). However,  the new stability criterion is based on  examining    the energy density functional for any arbitrary (permissible or impermissible) small variation above the background of the solitary wave solutions.  This criterion was used  without naming  in Derrick's article \cite{Derrick} about the nonexistent  of the stable non-vibrational  solitary wave   solutions of the  nonlinear Klein-Gordon field systems in  $3+1$ dimensions.  In general, a solitary wave solution which is stable according to the new criterion of the stability  can be called an energetically stable soliton solution.

Our main goal in this paper is to find a relativistic complex nonlinear field system that has just a single stable non-topological   solitary wave  solution (a single Q-ball solution). We expect, just for this single solitary  wave solution, the general   dynamical equations and other properties be    reduced to those versions  of a special type of the   standard well-known CNKG  systems. We have borrowed  this expectation  from the quantum field theory at which any standard  (nonlinear) Klein-Gordon (-like) system is used just to describe a special type of the known  particles.
 It should be noted that,  for  the known  CNKG systems \cite{Vak3,Vak4,Vak5,Vak6,Lee3,Scoleman,R1,R2,R3,R4,R5,R6,R7,Riazi2,MM3}, in general,  there are infinite solitary wave  solutions with different rest frequencies $\omega_{o}$, but this new system (which we call it extended CNKG system \cite{PH1,PH2,PH3}) has just a single solitary wave  solution with a specific rest frequency $\omega_{s}$ for which the general dynamical equations are reduced to the same standard CNKG versions, as  we expected. In other words, the simple CNKG system is a special case of the general extended CNKG system which is obtained just for the single solitary wave  solution.
 Furthermore, we expect  it to be a stable solution  according to the  new stability  criterion  which has been introduced in this paper;  that is, we expect its energy to be the minimum among the other (close) solutions. Nevertheless, for the single solitary wave solution,  introduced in this paper,  we will  show  that  it is also a stable solution  according to the Vakhitov-Kolokolov  and the quantum  mechanical stability   criteria   of the Q-balls.

To achieve these   goals, we add a new proper  term to the original CNKG Lagrangian density  in such away that  it  and all its derivatives will be equal to zero simultaneously just for the  single solitary wave  solution. This  new proper  additional term behaves  like  a  massless  spook\footnote{We chose  the word "\emph{spook}" in order to avoid any confusion with words such as   "\emph{ghost}" and "\emph{phantom}", which have their own particular meanings in the literature.}  which   surrounds the particle and resists any arbitrary deformations. There are some  parameters $A_{i}$'s and $B_{i}$'s ($i=1,2,3$) in the new additional term,   whose larger values result in  more stability of the single solitary wave  solution.  In other words,  the larger the values the greater will be the increase in the total energy for any arbitrary small variation above the background of the single solitary wave solution. The additional term, just makes the single solitary wave solution stable and does  not appear in any
of the observable, meaning that,  it acts  like a stability catalyser.
In fact, this model shows how we can have a nonlinear field system with a single stable non-topological solitary wave solution as a rigid particle,  for which the dominant dynamical equations are a special type of the standard CNKG equations.

Furthermore, it is necessary to say that   the most important motivation for writing this paper is the existence of several concerns about the fundamental understanding of particles in quantum  theory.
In general, quantum mechanics and quantum field theory only describe the probabilistic   behavior of tiny particles under the influence of other particles and environmental factors.
However, despite the undeniable successes of quantum theory, our knowledge of the fundamental particles is still scarce and there are still many unanswered questions about them.
For example, why are there few fundamental particles with specific masses, charges and other specific properties in nature? Why is the Planck  constant $\hbar$ the same for all particles  in the universe? Why are some fundamental particles,  such as muons, strange quarks,  essentially unstable?
Here and in the upcoming works, we do not claim that our models answer all these questions and describe real particles properly. But, trying to answer these questions has motivated us to develop a series of new mathematical tools first, and this article is an important step in this direction.

This paper has been organized as follows:   Basic equations and general properties of the CNKG systems with their  solitary wave solutions (Q-balls) are first considered. In the next section,   a new self-interaction potential and the corresponding localized wave solutions will be considered in detail, together with a stability analysis. In section IV, we will show  how to build an extended  CNKG system with a single stable solitary wave  solution  for which the general   dynamical equations are reduced  to  a  standard CNKG equation.   In section V,  the stability of the single soliton solution against any arbitrary small deformations will be studied according to the new criterion. In section VI, we provide a brief discussion about  the collisions of the single solitary wave  solutions with each other.   The last section is devoted to summary and conclusions. It should be note  that, this work is  in line with \cite{PH1,PH2}, but with more details and accuracy.

\section{ complex nonlinear  Klein-Gordon (CNKG) systems  }\label{sec2}

The complex nonlinear  Klein-Gordon (CNKG) systems in $1+1$ dimensions can be  introduced by the following  relativistic   Lagrangian-density:
 \begin{equation} \label{lag}
{\cal L}_{o}= \partial_\mu \phi^*
\partial^\mu \phi -V(|\phi|),
 \end{equation}
in which $\phi$ is a complex scalar field and $V(|\phi |)$ is the self-interaction potential, which depends only on the modulus of the scalar field. Using the least action principle, the dynamical
equation for the evolution of $\phi$ can be obtained as follows:
\begin{equation} \label{eq}
 \Box \phi =\frac{\partial^2\phi}{\partial
  t^2}- \frac{\partial^2\phi}{\partial
    x^2}=-\frac{\partial V}{\partial
  \phi^*}=-\frac{1}{2}V'(|\phi|)\frac{\phi}{|\phi|}.
  \end{equation}
  Note that we have used natural units $\hbar=c=1$ in this paper. For further applications, it is better to use  polar fields $R(x,t)$ and $\theta(x,t)$ as defined by
\begin{equation} \label{polar}
 \phi(x,t)= R(x,t)\exp[i\theta(x,t)].
\end{equation}
In terms of  polar fields, the Lagrangian-density (\ref{lag}) and related field equation (\ref{eq}) are reduced  respectively to
\begin{equation} \label{Lag2}
{\cal L}_{o}=(\partial^\mu R\partial_\mu R) +R^{2}(\partial^\mu\theta\partial_\mu\theta)-V(R),
\end{equation}
and
\begin{eqnarray} \label{e25}
&&\Box R-R(\partial^\mu\theta\partial_\mu\theta)=-\frac{1}{2}\frac{dV}{dR}, \\ \label{e252}&&
\partial_{\mu}(R^2\partial^{\mu}\theta)=2R(\partial_{\mu}R\partial^{\mu}\theta)+R^{2}(\partial^\mu\partial_\mu\theta)=0.
\end{eqnarray}
The related Hamiltonian (energy) density  is obtained via the Noether's theorem:
\begin{eqnarray} \label{TE}
&&T^{00}=\varepsilon(x,t)=\dot{\phi}\dot{\phi}^{*}+\acute{\phi}\acute{\phi}^{*}+V(|\phi |) \nonumber\\&&
=\dot{R}^{2}+\acute{R}^{2}+R^2(\dot{\theta}^{2}+\acute{\theta}^{2})+V(R),
\end{eqnarray}
in which dot and prime denote differentiation with respect to $t$ and $x$, respectively. For such systems, it is possible to have some  traveling solitary wave  solutions (Q-balls) as follows:
\begin{equation} \label{So}
R(x,t)=R(\gamma(x-vt)),\quad  \quad\theta(x,t)=k_{\mu}x^{\mu}=\omega t-kx,
\end{equation}
provided
\begin{equation} \label{pro}
k=\omega v,
\end{equation}
 and
 \begin{equation} \label{Re}
 \Box R=-\dfrac{d^2 R}{d\widetilde{x}^2}=-\frac{1}{2}\frac{dV}{dR}+\omega_{o}^{2}R,
 \end{equation}
where  $\gamma=1/\sqrt{1-v^2}$ and $\widetilde{x}=\gamma(x-vt)$. Note that $k^{\mu}\equiv(\omega,k)$ is a  $1+1$ dimensional   vector and  $\partial^\mu\theta\partial_\mu\theta=k_{\mu}k^{\mu}=\omega_{o}^{2}$ is a constant scalar.

 If we multiply equation (\ref{Re}) by $\frac{dR}{d\widetilde{x}}$ and integrate, it yields
\begin{equation} \label{St}
\left(\frac{dR(\widetilde{x})}{d\widetilde{x}}\right)^{2}+\omega_{o}^{2}R^2=V(R)+C,
\end{equation}
where $C$ is an integration constant. This constant is expected to vanish for a localized solitary wave solution. This equation can be easily solved for $R$, once the potential $V (R)$ is known:
\begin{equation} \label{ints}
\widetilde{x}-x_{o}=\pm\int\dfrac{dR}{\sqrt{V(R)-\omega_{o}^{2}R^2}},
\end{equation}
  In general, by using  equation (\ref{Re}), it is easy to see that there are different non-topological solutions for  $R(\widetilde{x})$ with different values of $\omega_{o}$. The topological complex kink and anti-kink solutions can also exist when $\omega_{o}=0$ and $V(R)$ has more than two vacuum points \cite{MM3}.

In the framework of special relativity, it is clear that the total energy of a solitary wave solution which represents the total relativistic energy of a particle should read
\begin{equation} \label{To}
 E=\int_{-\infty}^{+\infty}\varepsilon(x,t) dx=\gamma E_{o},
\end{equation}
in which
\begin{equation} \label{Tor}
 E_{o}=\int_{-\infty}^{+\infty}[\acute{R}^{2}+R^2\dot{\theta}^{2}+V(R)]dx=\int_{-\infty}^{+\infty}[\acute{R}^{2}+R^2\omega_{o}^{2}+V(R)]dx,
\end{equation}
is  the rest energy of a solitary wave  solution.

The Lagrangian-density (\ref{lag}) is invariant under global $U(1)$ symmetry. If we  link this
symmetry with electromagnetism, then according to Noether's theorem, the following electrical current
density is conserved:
\begin{equation} \label{cur}
j^\mu\equiv i (\phi\partial^\mu \phi^*-\phi^*\partial^\mu \phi)=2R^{2}\partial^{\mu}\theta,\quad  \partial_\mu j^\mu=0.
\end{equation}
 The corresponding charge is
\begin{equation} \label{Bar}
Q=\int_{-\infty}^{+\infty}j^0 dx=\int_{-\infty}^{+\infty}i(\phi\dot{\phi}^*-\phi^*\dot{\phi})dx,
\end{equation}
which is a constant of motion.

\section{Stability considerations; an example}\label{sec3}

\begin{figure}[ht!]
  \centering
  \includegraphics[width=110mm]{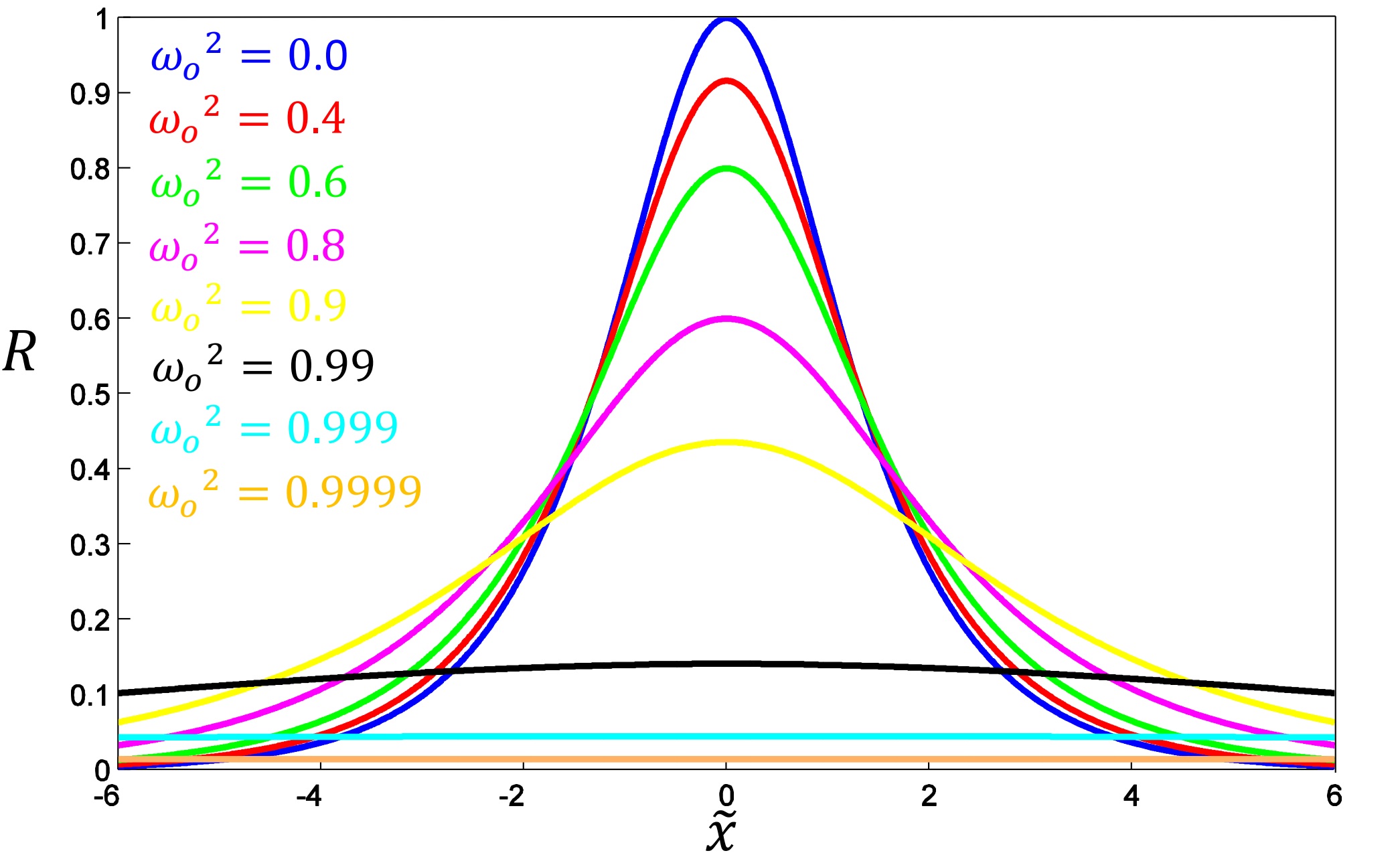}
  \caption{ Related to different $\omega_{o}^2$'s, there are different solutions for $R(\widetilde{x})$.} \label{DR}
\end{figure}

Based on what was done and introduced  in \cite{R5}, the following  potential is used  for simplicity in line with the purposes of this article:
\begin{equation} \label{po}
V(R)=R^2-R^4.
\end{equation}
With this  potential (\ref{po}), it can be shown that the integral ($\ref{ints}$) can be easily performed, yielding the following solutions for $\omega_{-}=0\leqslant \omega_{o}^2 < 1=\omega_{+}$:
\begin{equation} \label{S}
R(\widetilde{x})=\omega'\sech(\omega'\widetilde{x}).
\end{equation}
where $\omega'=\sqrt{1-\omega_{o}^2}$ is called the complementary   frequency \cite{R5}.
Accordingly, there are infinite solitary wave  solutions (\ref{S}), which can be identified with different rest frequencies $\omega_{o}$ (see Fig.~\ref{DR}).
Using Eqs.~(\ref{Tor}) and (\ref{Bar})  for  profile  functions (\ref{S}), then one can obtain the  rest  energies and  total charges:
\begin{equation} \label{mjh}
E_{o}=\frac{4\omega'}{3}(1+2\omega_{o}^2),\quad\quad\quad Q=4\omega_{o}\omega'.
\end{equation}

\begin{figure}[ht!]
  \centering
  \includegraphics[width=110mm]{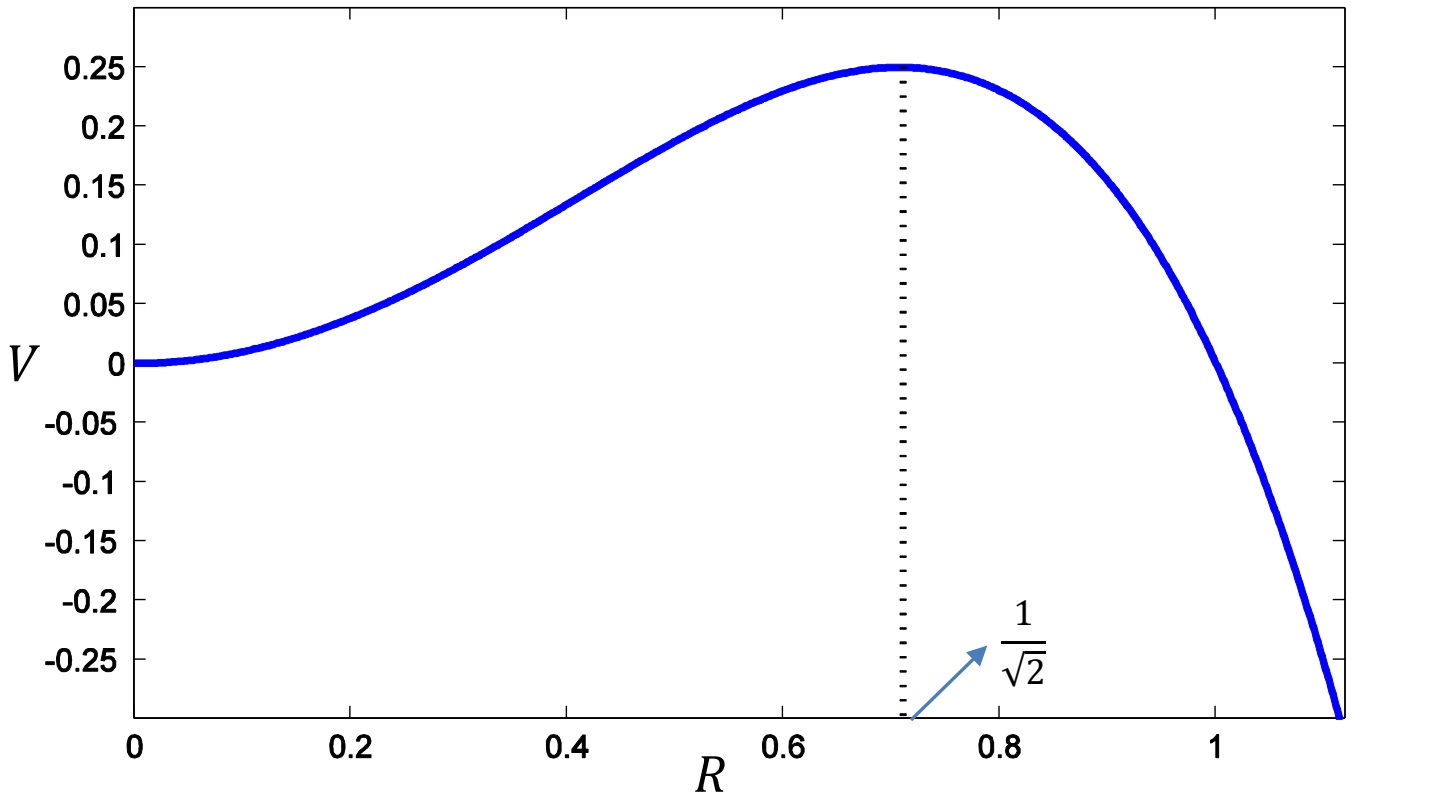}
  \caption{The field potential (\ref{po}) versus $R$} \label{pot}
\end{figure}

Traditionally there are two types  of criteria  for the stability of the Q-balls.
First, the quantum mechanical criterion \cite{Vak6,Lee3}, which  specifies    that  if the ratio between the rest energy and the charge is less than $\omega_{+}$ (i.e. $E_{o}/Q<\omega_{+}$) for a Q-ball solution, it can not decay to the free scalar particle quanta  with a specific  rest mass equal to $\omega_{+}$. Second, the classical  criterion \cite{Vak777,Vak3,Vak4,Vak5,Vak6},  which is based on  examining   the permissible small oscillating  perturbations  above the background of the Q-balls (not any arbitrary small deformations), it says that a Q-ball is stable if $\frac{dQ}{d\omega_{o}}<0$. Now, if these stability criteria  are used for the  system (\ref{po}) with the Q-ball solutions (\ref{S}), then it is easy to show that the Q-balls (\ref{S}) for which $\frac{1}{2}<\omega_{o}\leq 1$ ($\frac{1}{\sqrt{2}}=\omega_{c}<\omega_{o} \leq 1$) are quantum mechanically (classically) stable. Note that, for the case $\omega_{c}=\frac{1}{\sqrt{2}}$, the maximum value  of the related module function is  $R_{\textrm{max}}=R_{\omega_{c}}(0)=\frac{1}{\sqrt{2}}$ which is exactly the same  turning point  of the potential (\ref{po}) (see Fig.~\ref{pot}). In general, for any arbitrary  solution (\ref{S}),  it is obvious that the solutions for which $R_{\textrm{max}}>\frac{1}{\sqrt{2}}$ are essentially unstable. There is another type of  stability   called  fission stability. A Q-ball which does not  fulfill the  requirement of  the fission stability, it naturally   decays into two or more smaller Q-balls  with some release of energy. In general, it was shown that the condition of the classical  stability of the  Q-balls is identical to   the condition of  fission stability \cite{Vak6}. Therefore, the Q-balls  for which $\frac{dQ}{d\omega_{o}}<0$ are stable against fission too.


\begin{figure}[ht!]
  \centering
  \includegraphics[width=100mm]{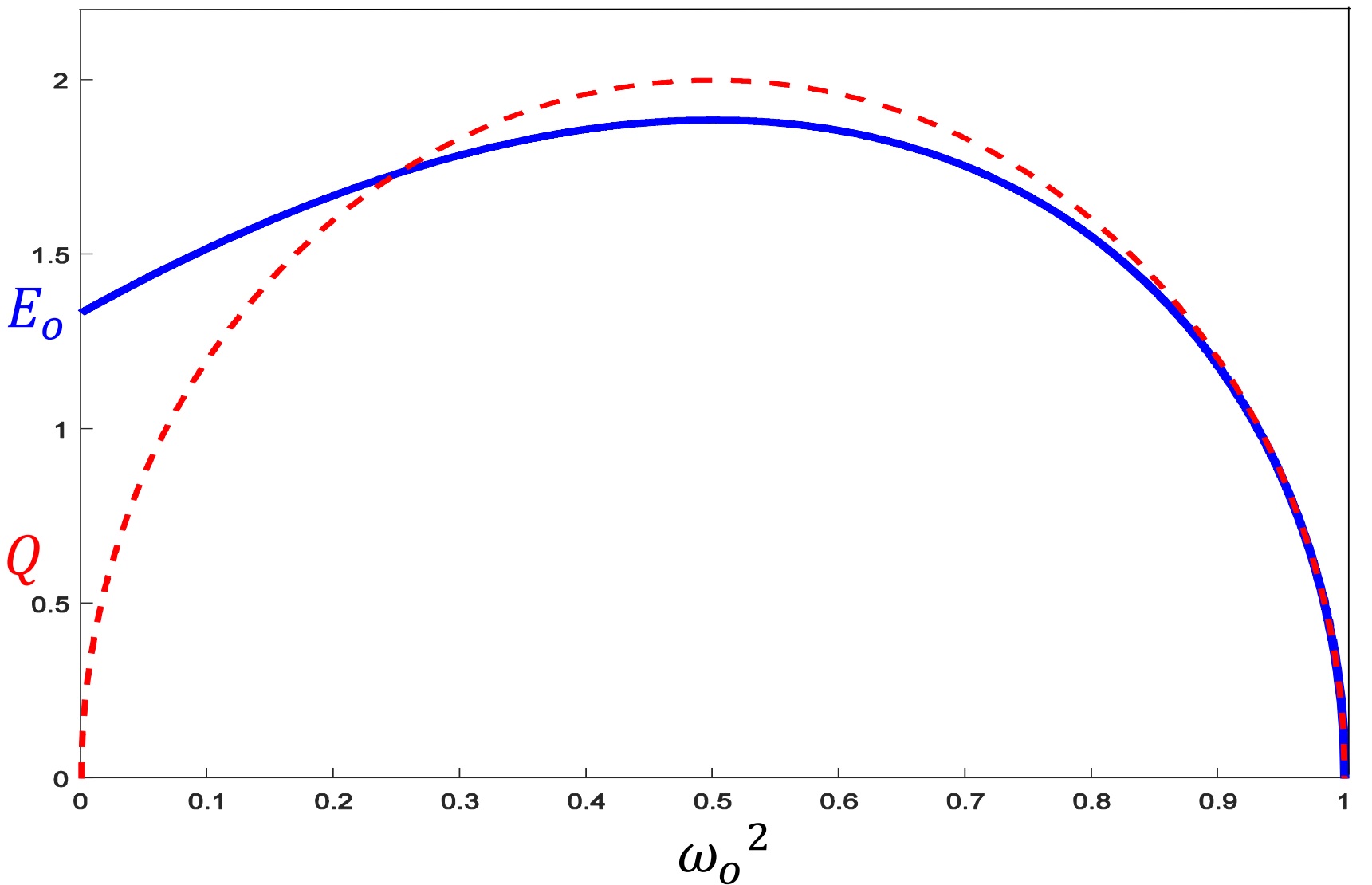}
  \caption{The rest energy (solid blue curve)  and  the total charge (dashed red curve) versus the rest  frequency square for different solitary wave  solutions. } \label{E}
\end{figure}

However, in this paper we use another rigorous criterion for the stability (i.e. the energetically stability criterion \cite{Derrick,PH1,PH2}).
The rest energy of  an  energetically stable solitary wave solution  is at the minimum among the other (close) solutions. In fact, if the  total energy always increases for any arbitrary (permissible or impermissible) deformation above the background of a special solitary wave solution which is at rest, it is  an energetically  stable solution.  Based on this criterion,  it  is easy to show that  there is no  stable solitary wave  solution for the CNKG systems in $1+1$ dimensions at all. For example, for a Q-ball solution, an arbitrary deformation (variation) can be constructed as follows: according to Eq.~(\ref{Tor}), let us fix the function $R(x)$ and set $\acute{\theta}=0$, then any small  variation in $\dot{\theta}^{2}$ with $\delta \dot{\theta}^{2}<0$,  yields a small reduction in the related  total  energy (\ref{Tor}).  Therefore, none of the   solitary wave  solutions (\ref{S}) are  energetically stable at all. The same argument applies  for  the possible non-topological solutions of the real nonlinear KG systems in $1+1$ dimensions.

Moreover,  based on the newly defined stability  criterion, a primary condition for a special solitary wave  solution (\ref{S}) to be a soliton,  is that
 its rest energy must be at the minimum among the other (close) solitary wave  solutions with different rest frequencies. According to Fig.~\ref{E}, there is not any solitary wave  solution with the minimum rest energy except $\omega_{o}=0$ and $\omega_{o}^2=1$. The case $\omega_{o}=0$ is a minimum, but for which $R_{\textrm{max}}=1>\frac{1}{\sqrt{2}}$, i.e. it is not essentially a stable object.   The case $\omega_{o}^2=1$ is the trivial  minimum according to the vacuum   $R(x)= 0$.

\section{An extended CNKG model with a single stable Q-ball  solution}\label{sec4}

In general,  there  is an  unwritten  postulate  in the quantum field  theory that states   any  standard Lagrangian density and the related dynamical equations of motion  are used just for a special type of the tiny  particles. Namely, the Dirac Lagrangian density with some specific inputs (electrons mass and charge) is used just for electrons and positrons, but the same Dirac Lagrangian densities with other specific inputs are used for other known particles like muons and neutrinos. Another example is a special version of the  complex nonlinear   Klein-Gordon Lagrangian densities (a complex $\phi^4$ system) which is used just to theoretically introduce the Higgs particles.  Moreover, in the standard relativistic   quantum field  theory, the Lagrangian densities, which are  used  for the  scalar fields  are usually  (complex) (nonlinear) Klein-Gordon types.

In light of the above, first we postulate that for the soliton solutions (as the particle-like objects) of the classical scalar field  models,   the   dominant Lagrangian densities (or the dominant dynamical equations)  should   be  the  same standard  (complex) (nonlinear) Klein-Gordon types.
Second, if a classical Lagrangian density  leads to more than one soliton solution with the same standard dominant dynamical equations, we postulate  that it  is not a physical case.  For example, the pervious Lagrangian density  (\ref{Lag2}) with the potential (\ref{po}) is not  a physical case, because it leads to infinite soliton-like solutions (\ref{S})  according to infinite particle-like objects for which the dominant Lagrangian density (\ref{Lag2}) or the dominant dynamical equations  (5) and (6) are  the same. Note that, we used   ''\emph{soliton-like}'' instead of  "\emph{soliton}", because, according to the new criterion of the stability, essentially none of them (\ref{S}) are energetically   stable   soliton solutions.
 So far, according to the new criterion of the stability, no classical field system has been introduced that leads to a non-topological stable soliton solution.
However, we will  introduce  a new field system in the following.
In fact, the topological property for many soliton solutions  is considered to guarantee the stability automatically and inevitably.
 The topological property of a  soliton solution imposes difficult conditions for constructing multi-particle solutions.  But with non-topological solutions, they simply result in multi-particle solutions just by adding them when they are sufficiently far apart.


To meet all these requirements, one can assume that there is  an    extended complex nonlinear Klein-Gordon (CNKG) Lagrangian density  \cite{PH1,PH2,PH3},  which is reduced to a simple standard CNKG form  just for a special   soliton solution.
Briefly, we are going to consider the possibility of   the existence of a new extended CNKG system with a single stable solitary wave  solution, according to the new criterion of the stability,  for which the  complicated  general    dynamical equations  are reduced  to the same standard CNKG versions  (5) and (6) as we expected.
Note that, unlike the quantum field	theory, the particle concept in the classical field theory is completely objective (i.e. a stable localized energy density in the  space which can move in any arbitrary direction).




To make it more objective, we can imagine that a stable solitary wave solution of an unknown relativistic field	system with a specific rest frequency $\omega_{s}=0.8$ exists in following form:
\begin{eqnarray} \label{SS}
&&\phi_{s}(x,t)=R_{s}(\gamma(x-vt))\exp{(ik_{\mu}x^{\mu})}=R_{s}(\widetilde{x})\exp{(i\gamma\omega_{s}(t-vx))}\nonumber\\&&
\quad\quad\quad\quad=\omega_{s}'\sech (\omega_{s}'\widetilde{x})\exp{(i\omega_{s}\widetilde{t}~)},
\end{eqnarray}
in which  $\widetilde{t}=\gamma(t-vx)$ and $\omega_{s}'=0.6$.
In other words, it is considered to be one of the  Q-balls (\ref{S}) which is quantum mechanically and classically stable.   We can consider this special solitary wave  solution (SSWS) (\ref{SS}) like a  detected stable tiny   particle in the laboratory  for which  we suppose that  the  dominant dynamical equations of motion or the dominant Lagrangian density be the  same standard versions of the   CNKG system which were introduced  in the previous sections.  In other words,   we assume that the dynamical equations  of the system have  a  general complicated  form, and just for the SSWS (\ref{SS}) are reduced  to the same simple  standard CNKG  forms (\ref{e25}) and (\ref{e252}).

Accordingly, one should  consider a new Lagrangian density   in the following  form:
\begin{equation} \label{lag2}
{\cal L}= {\cal L}_{o}+F=[(\partial^\mu R\partial_\mu R) +R^{2}(\partial^\mu\theta\partial_\mu\theta)-V(R)]+F,
 \end{equation}
in which ${\cal L}_{o}$ is the same original CNKG Lagrangian density (\ref{Lag2}) for which   the    SSWS (\ref{SS}) is  one of its  solutions.
According to the standard classical relativistic field theory, the  Lagrangian densities are considered to be functions of the fields themselves  and their first derivatives. In addition, since the Lagrangian densities must be scalar functionals,  they should be functions of the possible allowed scalars. Along with the scalar fields $R$ and $\theta$,  the  other  basic (simplest) allowed scalars in our model, which are made via  the different possible contractions  of  the first derivatives of the scalar fields, are $\partial_{\mu}R\partial^{\mu}R$, $\partial_{\mu}\theta\partial^{\mu}\theta$ and $\partial_{\mu}R\partial^{\mu}\theta$. Accordingly, we conclude that the new additional functional $F$ (which is a  part of the new lagrangian density) should  be function  of all possible allowed scalars $R$, $\theta$, $\partial_{\mu}R\partial^{\mu}R$, $\partial_{\mu}\theta\partial^{\mu}\theta$ and $\partial_{\mu}R\partial^{\mu}\theta$. To ensure that the electrical charge conservation is satisfied again, the additional term $F$ must not be function of the phase field $\theta$. The new dynamical  equations of motion for this new extended Lagrangian density (\ref{lag2}) are
\begin{eqnarray} \label{geq}
&&\Box R-R(\partial^\mu\theta\partial_\mu\theta)+\frac{1}{2}\frac{dV}{dR}+\frac{1}{2}\left[ \frac{\partial}{\partial x^{\mu}}\left(\frac{\partial F }{\partial (\partial_{\mu}R)}\right)-\left(\frac{\partial F}{\partial R}\right)  \right]=0\\ \label{geq2}&&  \partial_{\mu}(R^2\partial^{\mu}\theta)+\frac{1}{2} \left[ \frac{\partial}{\partial x^{\mu}}\left(\frac{\partial F }{\partial (\partial_{\mu}\theta)}\right)  \right]=0.
\end{eqnarray}
Also, the new  energy density function is
\begin{eqnarray} \label{e111}
&&\varepsilon(x,t)=\left[\dot{R}^{2}+\acute{R}^{2}+R^2(\dot{\theta}^{2}+\acute{\theta}^{2})+V(R)\right]+\left[\dot{R}\frac{\partial F}{\partial \dot{R}}+\dot{\theta}\frac{\partial F}{\partial \dot{\theta}}\right].
\end{eqnarray}
For the SSWS (\ref{SS}), as we indicated   before, we expect all Eqs. (\ref{lag2}), (\ref{geq}), (\ref{geq2}) and (\ref{e111}) to be reduced  to the same original versions  (\ref{lag}), (\ref{e25}), (\ref{e252}) and (\ref{TE}) respectively. In other words, we expect all additional terms $F$, $\frac{\partial}{\partial x^{\mu}}\left(\frac{\partial F }{\partial (\partial_{\mu}R)}\right)$, $\frac{\partial}{\partial x^{\mu}}\left(\frac{\partial F }{\partial (\partial_{\mu}\theta)}\right)$, $\frac{\partial F}{\partial R}$, $\frac{\partial F}{\partial \dot{R}}$ and $\frac{\partial F}{\partial \dot{\theta}}$ to be zero just for the SSWS (\ref{SS}). It means that just  for the SSWS (\ref{SS}), the general  equations of motion  (23) and (24) turn to  the same standard   equations (4) and (5) respectively.


 Therefore,  first, we  find    the standard CNKG  Lagrangian density   ${\cal L}_{o}$  as  the dominant Lagrangian density for the  SSWS (\ref{SS}). Second, we try to find  the proper additional  term $F$  in  such a way that ${\cal L}_{o}$ to be the  dominant Lagrangian density just for  the SSWS (\ref{SS}), meaning that, the additional term $F$ and its other derivatives simultaneously turn to zero just for the SSWS (\ref{SS}). And third, we expect this additional term to guarantee the energetically  stability of the  SSWS (\ref{SS}), meaning that  the rest energy of the SSWS (\ref{SS}) be  at a minimum among the other (close) solutions of the new system (\ref{lag2}).
Note that, for this new relativistic field system (\ref{lag2}), there is just a single solitary wave  solution (Q-ball) (\ref{SS}) with a specific rest frequency $\omega_{s}=0.8$, that is,  the other Q-balls (\ref{S}) of the original system (\ref{lag}) with different rest frequencies are not the solutions of this new system (\ref{lag2}) anymore.

Since $F$ and all its  derivatives  must be zero for the SSWS (\ref{SS}), one can conclude that  it should   be a function of powers of $\mathbb{S}_{i}$'s, where $\mathbb{S}_{i}$'s are introduced as the possible
independent  scalars which are zero simultaneously for the SSWS (\ref{SS}). As mentioned earlier, in general, $F$ must be a function of the allowed  scalars,  on the other hand, $F$ is considered  to be a function of the powers of the $\mathbb{S}_{i}$'s, thus $\mathbb{S}_{i}$'s must be functions of the allowed  scalars as well.  Therefore, there are just three basic independent combinations of the   allowed  scalars, which would be zero for the SSWS (\ref{SS}) simultaneously as follows:
\begin{eqnarray} \label{sc2}
  &&\mathbb{S}_{1}=\partial_{\mu}\theta\partial^{\mu}\theta-\omega_{s}^2,\\ \label{sc21} &&
 \mathbb{S}_{2}=\partial_{\mu}R\partial^{\mu}R+V(R)-\omega_{s}^2 R^2, \\ \label{sc22} &&
 \mathbb{S}_{3}=\partial_{\mu}R\partial^{\mu}\theta.
\end{eqnarray}
It is straightforward  to show that these special scalars all are equal to zero for the SSWS (\ref{SS}).
For simplicity’s sake, if one  considers  $F$ as a function of  n'th power of  $\mathbb{S}_{i}$'s, i.e. $F=F(\mathbb{S}_{1}^{n},\mathbb{S}_{2}^{n},\mathbb{S}_{3}^{n})$, it yields
\begin{eqnarray} \label{sc3}
 && \frac{\partial}{\partial x^{\mu}}\left(\frac{\partial F }{\partial (\partial_{\mu}R)}\right)=\sum_{i=1}^{3} \left[n(n-1)\mathbb{S}_{i}^{(n-2)}\frac{\partial \mathbb{S}_{i}}{\partial x^{\mu}}\frac{\partial \mathbb{S}_{i} }{\partial (\partial_{\mu}R)} \frac{\partial F}{\partial Z_{i}}+n\mathbb{S}_{i}^{(n-1)}\frac{\partial}{\partial x^{\mu}}\left(\frac{\partial \mathbb{S}_{i} }{\partial (\partial_{\mu}R)} \frac{\partial F}{\partial Z_{i}}\right) \right] \nonumber  \\&&
\frac{\partial F}{\partial R}=\sum_{i=1}^{3}\left[n\mathbb{S}_{i}^{(n-1)}\frac{\partial \mathbb{S}_{i}}{\partial R}\frac{\partial F}{\partial Z_{i}}\right]\nonumber\\&&
\frac{\partial}{\partial x^{\mu}}\left(\frac{\partial F }{\partial (\partial_{\mu}\theta)}\right)=\sum_{i=1}^{3} \left[n(n-1)\mathbb{S}_{i}^{(n-2)}\frac{\partial \mathbb{S}_{i}}{\partial x^{\mu}}\frac{\partial \mathbb{S}_{i} }{\partial (\partial_{\mu}\theta)} \frac{\partial F}{\partial Z_{i}}+n\mathbb{S}_{i}^{(n-1)}\frac{\partial}{\partial x^{\mu}}\left(\frac{\partial \mathbb{S}_{i} }{\partial (\partial_{\mu}\theta)} \frac{\partial F}{\partial Z_{i}}\right) \right] \nonumber.
\end{eqnarray}
where $Z_{i}=\mathbb{S}_{i}^n$. It is easy to understand for $n\geq3$, all these relations would be zero for the SSWS (\ref{SS}) as we expected. Accordingly, one can show that   the   general form of the functional  $F$ which satisfies  all required   constraints, can be introduced by a  series:
\begin{eqnarray} \label{sf0}
 F=\sum_{n_{3}=0}^{\infty}\sum_{n_{2}=0}^{\infty}\sum_{n_{1}=0}^{\infty} a({n_{1},n_{2},n_{3}})\mathbb{S}_{1}^{n_{1}}\mathbb{S}_{2}^{n_{2}}\mathbb{S}_{3}^{n_{3}},
\end{eqnarray}
provided $(n_{1}+n_{2}+n_{3})\geq3$.    Note that,    coefficients $a({n_{1},n_{2},n_{3}})$  are also arbitrary well-defined functional scalars, i.e. they can be again functions of all possible allowed  scalars   $R$, $\partial_{\mu}R\partial^{\mu}R$, $\partial_{\mu}\theta\partial^{\mu}\theta$ and $\partial_{\mu}R\partial^{\mu}\theta$ (except $\theta$).


The  stability  conditions  impose serious  constraints  on  function $F$  which causes series (\ref{sf0}) to be reduce to  special  formats. However, again there are many choices which can  lead to a stable SSWS (\ref{SS}). Among them,  one can consider the additional term in the following form:
 \begin{equation} \label{F}
 F=\sum_{i=1}^{3} A_{i}f(Z_{i}),
 \end{equation}
where $Z_{i}=B_{i}{\cal K}_{i}^n$ for which $n$ is any arbitrary odd number larger than $1$ (i.e. $n=3,5,7,\cdots$),  $f(Z_{i})$ is  any arbitrary odd $\sinh$-like  function whose  odd derivatives   are all  non-negative at $Z_{i}=0$  (for example $f=Z_{i}$ \cite{PH1} or $f=Z_{i}^3$),  $A_{i}$'s and $B_{i}$'s ($i=1,2,3$) are just some positive constants, and  functionals ${\cal K}_{i}$'s  are three independent linear combinations of $\mathbb{S}_{i}$'s as follows:
 \begin{eqnarray} \label{ek}
  &&{\cal K}_{1}= R^2 \mathbb{S}_{1},\\ \label{ek1}&&
{\cal K}_{2}= R^2 \mathbb{S}_{1}+\mathbb{S}_{2}, \\ \label{ek3}&&
{\cal K}_{3}= R^2 \mathbb{S}_{1}+\mathbb{S}_{2}+2 R \mathbb{S}_{3},
\end{eqnarray}
 It is  obvious that ${\cal K}_{1}$, ${\cal K}_{2}$ and ${\cal K}_{3}$ are all zero just for the SSWS (\ref{SS}) with the rest frequency $\omega_{s}=\pm \frac{4}{5}=\pm 0.8$. Note that, these special linear combinations of the $\mathbb{S}_{i}$'s in Eqs.~(\ref{ek})-(\ref{ek3})   are introduced just  in  line  with the objectives of this paper and  are not unique.  One can use other  combinations to obtain different systems with different properties.

The energy-density (\ref{e111}) that belongs to the new extended  Lagrangian-density (\ref{lag2}), for this  special choice (\ref{F}),  turns to
\begin{eqnarray} \label{MTEi}
&&\varepsilon(x,t)=\left[\dot{R}^{2}+\acute{R}^{2}+R^2(\dot{\theta}^{2}+\acute{\theta}^{2})+V(R)\right]+\nonumber\\&&\sum\limits_{i=1}^{3} \left[nA_{i}B_{i}
C_{i}{\cal K}_{i}^{n-1}f_{i}'-A_{i}f(Z_{i})\right]=\varepsilon_{o}+\varepsilon_{1}+\varepsilon_{2}+\varepsilon_{3},
\end{eqnarray}
where $f_{i}'=\dfrac{df(Z_{i})}{dZ_{i}}$,  and
\begin{equation}\label{cof}
C_{i}=\dfrac{\partial{\cal K}_{i}}{\partial \dot{\theta}}\dot{\theta}+\dfrac{\partial{\cal K}_{i}}{\partial \dot{R}}\dot{R}=
\begin{cases}
\quad\quad 2 R^2 \dot{\theta}^{2} & \text{i=1}
\\
2(\dot{R}^{2}+ R^2 \dot{\theta}^2) & \text{i=2}
\\
2(\dot{R}+ R \dot{\theta})^{2}
 & \text{i=3}.
\end{cases}
 \end{equation}
Note that, $C_{i}$'s are positive definite and this property will be used in the further  conclusions. In fact, this main property originates  from the proper combination of the $\mathbb{S}_{i}$'s in Eqs.~(\ref{ek})-(\ref{ek3})  to introduce special functionals  ${\cal K}_{1}$, ${\cal K}_{2}$ and ${\cal K}_{3}$.

 Since $f(Z_{i})$ is considered  an odd $\sinh$-like function, hence  it can be shown  generally by a  convergent Maclaurin's series
\begin{equation} \label{Se}
f(Z_{i})=\sum\limits_{j=0} ^{\infty}a_{j}Z_{i}^{2j+1}=\sum\limits_{j=0}^{\infty}a_{j}B_{i}^{2j+1}{\cal K}_{i}^{2nj+n},
\end{equation}
where $a_{j}$'s are all non-negative.  It is easy to obtain $f_{i}'$ (as an even function) in a series format:
\begin{equation} \label{Sr}
f_{i}'=\dfrac{df(Z_{i})}{dZ_{i}}=\sum\limits_{j=0}^{\infty}a_{j}(2j+1)Z_{i}^{2j}=\sum\limits_{j=0}^{\infty}a_{j}(2j+1)B_{i}^{2j}{\cal K}_{i}^{2nj}.
\end{equation}
Now,  functions $\varepsilon_{i}$'s ($i=1,2,3$) in  Eq.~(\ref{MTEi}) can be expressed in the following series:
\begin{equation} \label{e11}
\varepsilon_{i}=\left[nA_{i}B_{i}
C_{i}{\cal K}_{i}^{n-1}f_{i}'-A_{i}f(Z_{i})\right]=A_{i}\sum\limits_{j=0}^{\infty}a_{j}B_{i}^{2j+1}{\cal K}_{i}^{2nj+n-1}D_{ij},
\end{equation}
where $D_{ij}=\left[nC_{i}(2j+1)-{\cal K}_{i} \right]$. Since $n$ is considered  as equal to a positive odd  number, the power of  ${\cal K}_{i}$'s (i.e. $2nj+n-1$) in the Eq.~(\ref{e11}) would be always even numbers. Moreover, since   $a_{j}\geqslant 0$, like a $\sinh$ function,   the terms $A_{i}a_{j}B_{j}^{2j+1}{\cal K}_{i}^{2nj+n-1}$  in Eq.~(\ref{e11}) will be   positive definite  and are always zero just for the SSWS (\ref{SS}). Now, from Eq.~(\ref{cof}), one can easily calculate  coefficients $D_{ij}=\left[nC_{i}(2j+1)-{\cal K}_{i} \right]$:
\begin{equation}\label{coff}
D_{ij}=
\begin{cases}
  R^2 [5\dot{\theta}^2+\acute{\theta}^2+\omega_{s}^2]+C_{1}(2jn+n-3) & \text{i=1}
\\
[5 R^2 \dot{\theta}^2+5\dot{R}^2+ R^2 \acute{\theta}^2+\acute{R}^2+U(R)]+C_{2}(2jn+n-3), & \text{i=2}
\\
[5( R \dot{\theta}+\dot{R})^2+( R \acute{\theta}+\acute{R})^2+U(R)]+C_{3}(2jn+n-3),
 & \text{i=3}
\end{cases}
 \end{equation}
where $U(R)=2\omega_{s}^2 R^2 - V(R)=R^4+\frac{7}{25} R^2$,  which   is a  non-negative function and bounded from below.     Therefore,  since  $n\geq 3$ and $C_{i}$'s  are positive  definite, we are sure that all terms in the above relations are positive definite which means that all terms of the series (\ref{e11}) would be positive definite. In other words, all $\varepsilon_{i}$'s ($i=1,2,3$) are positive definite functions which are zero just for the SSWS (\ref{SS}) and the vacuum ($R=0$) simultaneously.

Since ${\cal K}_{1}$, ${\cal K}_{2}$ and ${\cal K}_{3}$ (or equivalently $\mathbb{S}_{1}$, $\mathbb{S}_{2}$ and $\mathbb{S}_{3}$) are three independent scalars for two scalar fields $R$ and $\theta$, it is not possible to find a special  variation in  the SSWS (\ref{SS}) for which all of ${\cal K}_{i}$'s  do not change and stay zero simultaneously. In other words, just for the SSWS (\ref{SS}) (and the vacuum $R=0$),   all ${\cal K}_{i}$'s would be zero simultaneously and for other non-trivial solutions of the extended CNKG system (\ref{lag2}), at least one of the ${\cal K}_{i}$'s ($\mathbb{S}_{i}$'s) would be a non-zero function (see the Appendix A). Therefore, if constants $A_{i}$'s or $B_{i}$'s are considered to be large numbers, we expect for other   solutions of the new extended system (\ref{lag2}),  according to Eq.~(\ref{e11}), at least one of  $\varepsilon_{i}$'s would be a  large positive function, and then the related  rest energy would be  larger than SSWS  rest energy. Accordingly, we expect the rest energy of the SSWS (\ref{SS}) would  be at a minimum among the other solutions, except the ones which are very close to the vacuum state ($R=0$).

To summarize, the odd functions $f(Z_{i})$ for which the coefficients of the related Maclaurin's series (\ref{Se}) are all non-negative (i.e. $a_{j}\geqslant 0$),  are the proper functions to guarantee the stability of the  SSWS (\ref{SS}). In fact, for these special odd $\sinh$-like functions $f(Z_{i})$, the additional terms of the energy density function (\ref{MTEi}), i.e.  $\varepsilon_{1}$, $\varepsilon_{2}$ and $\varepsilon_{3}$, would be positive  definite functions and all are zero simultaneously just for the  SSWS (\ref{SS}). To prove that the SSWS (\ref{SS}) is genuinely   a stable object, we just  considered  functions $\varepsilon_{i}$'s ($i=1,2,3$)  but we did not consider  function $\varepsilon_{o}$! In the next section, we will show that theoretically and numerically for  systems with large enough values of $B_{i}$'s (or $A_{i}$'s), the influence of the function $\varepsilon_{o}$ in the   stability  property is small and negligible.

\section{stability for small deformations}\label{sec5}

In this section,  based on  the new criterion of the stability (energetically stability criterion),  we are going to study  the variations  of the  total energy above  the background of the  SSWS (\ref{SS}) for small variations.
 In general, the arbitrary  small variations for the non-moving SSWS (\ref{SS}) can be considered  as follows:
\begin{equation} \label{so1}
R(x,t)=R_{s}(x)+\delta R(x,t) \quad \textrm{and} \quad \theta(x,t)=\theta_{s}(t)+\delta \theta(x,t)=\omega_{s}t+\delta \theta(x,t),
\end{equation}
where $\delta R$ and $\delta \theta$ (small variations) are  any  small functions  of space-time. The subscript $s$  is referred to the special solution (\ref{SS}) for which $\omega_{s}^2=0.8$ and $R_{s}(x)=0.6\sech (0.6 x)$.
Now, if we insert  the deformed version of the non-moving  SSWS (\ref{so1}) in    $\varepsilon_{o}(x,t)$ and keep the terms up to the first order of  variations, then it yields
\begin{eqnarray} \label{so3}
&&\varepsilon_{o}(x,t)=\varepsilon_{os}(x)+\delta\varepsilon_{o}(x,t)\approx \left[\acute{R_{s}}^2  +R_{s}^2\omega_{s}^2+V(R_{s})\right]+\nonumber\\&&
2\left[ \acute{R_{s}} (\delta \acute{R})+R_{s}(\delta R)\omega_{s}^2+
   R_{s}^{2}\omega_{s}(\delta\dot{\theta})+\frac{1}{2}\frac{dV(R_{s})}{dR_{s}}(\delta R)\right].
\end{eqnarray}
Note that, for a non-moving SSWS (\ref{SS}), $\dot{R_{s}}=0$, $\acute{\theta_{s}}=0$ and $\dot{\theta_{s}}=\omega_{s}=\pm \sqrt{0.8}$. Therefore, $\delta\varepsilon_{o}$  can be considered as a linear  function of the first order of small variations $\delta R$, $\delta \acute{R}$ and $\delta \dot{\theta}$. It is obvious that $\delta\varepsilon_{o}$ is not necessarily a positive definite function for  arbitrary variations.
If one performs  the similar procedure   for $\varepsilon_{i}$'s, they lead to
\begin{eqnarray} \label{fvfv}
&&\varepsilon_{i}(x,t)=\varepsilon_{is}(x)+\delta\varepsilon_{i}(x,t)=\delta\varepsilon_{i}(x,t)\nonumber\\&&
=A_{i}\sum\limits_{j=0}^{\infty}a_{j}B_{i}^{2j+1}\left[(D_{ijs}+\delta D_{ij} )({\cal K}_{is}+\delta{\cal K}_{i})^{2nj+n-1}\right]
\end{eqnarray}
 Note that ${\cal K}_{i}$'s for the SSWS (\ref{SS}) would be zero (i.e. ${\cal K}_{is}=0$).  Now, for simplicity, if one sets $n=3$,  then
\begin{eqnarray} \label{ddf}
&&\delta\varepsilon_{i}(x,t)=
A_{i}\sum\limits_{j=0}^{\infty}a_{j}B_{i}^{2j+1}\left[(D_{ijs}+\delta D_{ij} )(\delta{\cal K}_{i})^{6j+2}\right]\approx A_{i}a_{0}B_{i} D_{i0s} (\delta{\cal K}_{i})^{2},
\end{eqnarray}
According to Eq.~(\ref{cof}), $D_{i0s}=3C_{is}-{\cal K}_{is}=3C_{is}=6  R_{s}^2 \omega_{s}^2$, then Eq.~(\ref{ddf}) is simplified  to
\begin{equation}\label{kbj}
\delta\varepsilon_{i}(x,t)\approx 6 A_{i}B_{i}a_{0}  R_{s}^2 \omega_{s}^2 (\delta{\cal K}_{i})^{2}\propto A_{i}B_{i}(\delta{\cal K}_{i})^{2}\geq 0,
\end{equation}
hence, for small variations $\delta\varepsilon_{i}$'s are all positive definite functions as we generally expected.

It is easy to show that $\delta{\cal K}_{i}$'s, similar to $\delta\varepsilon_{o}$,  are all linear functions  of  the first order of small variations. In fact, according to Eqs.~(\ref{ek})-(\ref{ek3}) and (\ref{sc2})-(\ref{sc22}) we can define three linear  functions $G_{1}$, $G_{2}$ and $G_{3}$ in terms of  small variations as follows:  $\delta{\cal K}_{1}=G_{1}(\delta \dot{\theta})=2\omega_{s}R_{s}^2\delta \dot{\theta}$, $\delta{\cal K}_{2}=G_{2}(\delta R,\delta \acute{R},\delta \dot{\theta})=G_{1}-2\acute{R_{s}} (\delta \acute{R})+(\frac{dV(R_{s})}{dR_{s}}-2\omega_{s}^2 R_{s})\delta R$ and $\delta{\cal K}_{3}=G_{3}(\delta R,\delta \acute{R},\delta\dot{R},\delta\acute{\theta},\delta \dot{\theta})=G_{2}+2 R_{s}(\omega_{s}\delta\dot{R}-\acute{R}_{s}\delta \acute{\theta}) $ respectively.
Hence, from Eq.~(\ref{kbj}), one can simplify conclude that $\delta\varepsilon_{i}$ ($i=1,2,3$) is  a linear function of the second order of small variations which is multiplied by   coefficient $A_{i}B_{i}$. In other words, we can define three linear functions $W_{1}$, $W_{2}$ and $W_{3}$ in such a way that $\delta\varepsilon_{1}=A_{1}B_{1}~W_{1}([\delta \dot{\theta}]^2)$, $\delta\varepsilon_{2}=A_{2}B_{2}~W_{2}([\delta R]^2,[\delta \acute{R}]^2,[\delta \dot{\theta}]^2, \delta R\delta \acute{R}, \delta R\delta\dot{\theta},\delta\acute{R}\delta\dot{\theta})$ and  $\delta\varepsilon_{3}=A_{3}B_{3}~W_{3}([\delta R]^2,[\delta \acute{R}]^2,[\delta \dot{R}]^2,[\delta \acute{\theta}]^2,$ $[\delta \dot{\theta}]^2, \delta R\delta \acute{R},\delta R\delta \dot{R}, \delta R\delta\acute{\theta},\cdots,\delta \acute{\theta}\delta\dot{\theta} )$. For small variations, it is obvious that
the magnitude of the first order of variations is larger than the magnitude of the second order of them (for example, $\delta R<(\delta R)^2$), hence, it is easy to understand that for small variations: $W_{i}<G_{i}$ or $W_{i}<\delta\varepsilon_{o}$ ($i=1,2,3$). But, if  constants $A_{i}$'s or $B_{i}$'s are considered to be large numbers, the comparison between $\delta\varepsilon_{i}=A_{i}B_{i}W_{i}$ and $G_{i}$ (or $\delta\varepsilon_{o}$) needs more considerations. For example, if one considers  $A_{i}=B_{i}=10^{20}$, then  for the  variations larger (smaller)  than $\delta R=10^{-10}$ we have $\delta R< A_{i}B_{i}(\delta R)^2$ ($\delta R> A_{i}B_{i}(\delta R)^2$), hence   the same argument goes  for the comparison between $\delta\varepsilon_{i}$'s and $G_{i}$'s or the comparison between $\delta\varepsilon_{i}$'s and $\delta\varepsilon_{o}$.

Accordingly, if constants $A_{i}$'s and $B_{i}$'s are not  large numbers,  it is obvious that $|\delta\varepsilon_{o}|<\sum_{i=1}^{3}\delta\varepsilon_{i}$ for all small deformations. But, if constants $A_{i}$'s and $B_{i}$'s are considered to be   large numbers,  $|\delta\varepsilon_{o}|$ just for too small variations  may be larger than $\sum_{i=1}^{3}\delta\varepsilon_{i}$, and then the variation of the total energy density may be negative, i.e. it may be  $\delta\varepsilon=\delta\varepsilon_{o}+\sum_{i=1}^{3} \delta\varepsilon_{i} < 0$. For such too small variations the stability conditions of the new criterion  may   not fulfilled; nevertheless, they are physically too small which can be ignored in stability considerations. In fact, these  too small variations  are a sign  of the fact that,  the dominant dynamical equations of motion for the SSWS (\ref{SS}) are the same standard original  CNKG equations (5) and (6).
 Therefore, like a chicken in the egg in which its  internal movements are confined   by the  egg shell, this SSWS (\ref{SS}) can have   some unimportant  internal deformations which are confined by the additional term $F$ in the new system (\ref{lag2}).

To summarize, if  we consider the extended CNKG systems with large  $A_{i}$'s or $B_{i}$'s,  $\delta\varepsilon$ would be always positive for all  significant physical variations ($\delta R$ and $\delta \theta$) and then  the stability of the SSWS would guaranteed   appreciably. Just for some unimportant  too small variations, it may be possible to  see the violation of the stability, but  the rest energy reduction  for these  variations are  so small  that they can be ignored physically. Although, the $A_{i}$'s and $B_{i}$'s
parameters can be taken as very large values, but they will not affect the dynamical equations and the other properties
of the SSWS (\ref{SS}). In other words, the additional  term $F$ (\ref{F}) in the new system (\ref{lag2}) with large values of parameters $B_{i}$'s (or $A_{i}$'s) behaves like a stability catalyser, but  does  not have any role in  the observables of the SSWS (\ref{SS}). In the following, we will introduce many  arbitrary variations and will show numerically  how considering  systems with large  $A_{i}$'s and $B_{i}$'s appreciably guarantees the stability of the SSWS (\ref{SS}).

From now on, according to Eq.~(\ref{Se}) and the pervious discussions,   let us   consider an  odd function in the following form:
\begin{equation} \label{examp}
f(Z_{i})=\sinh(Z_{i}),
 \end{equation}
where $Z_{i}=B_{i}{\cal K}_{i}^3$.  Therefore, the related extended  Lagrangian density is
\begin{equation} \label{LN}
 {\cal L}=\left[\partial^\mu R\partial_\mu R +R^{2}(\partial^\mu\theta\partial_\mu\theta)-V(R)\right]+\sum_{i=1}^{3} A_{i}\sinh(B_{i}
 {\cal K}_{i}^3)={\cal L}_{o}+{\cal L}_{1}+{\cal L}_{2}+{\cal L}_{3}.
 \end{equation}
The related equations of motion are
\begin{eqnarray} \label{e5}
&&\left[\Box R-R(\partial^\mu\theta\partial_\mu\theta)+\frac{1}{2}\frac{dV}{dR}\right]+\sum_{i=2}^{3}\left [\frac{3}{2}A_{i}B_{i} \dfrac{\partial}{\partial x^{\mu}}
\left({\cal K}_{i}^2\frac{\partial {\cal K}_{i}}{\partial(\partial_\mu R)}\cosh(B_{i}
 {\cal K}_{i}^3)\right)\right]-\nonumber\\&&
\sum_{i=2}^{3}\left[\frac{3}{2}A_{i}B_{i}
\left({\cal K}_{i}^2\frac{\partial {\cal K}_{i}}{\partial R}\cosh(B_{i}{\cal K}_{i}^3)\right)\right]=0,\\&&
\partial_{\mu}(R^2\partial^{\mu}\theta)+\sum_{i=1}^{2}\left [\frac{3}{2}A_{i}B_{i} \dfrac{\partial}{\partial x^{\mu}}
\left({\cal K}_{i}^2\frac{\partial {\cal K}_{i}}{\partial(\partial_\mu \theta)}\cosh(B_{i}
 {\cal K}_{i}^3)\right)\right]=0,
\end{eqnarray}
and the related energy density is
\begin{eqnarray} \label{MTE}
&&\varepsilon(x,t)=\left[\dot{R}^{2}+\acute{R}^{2}+R^2(\dot{\theta}^{2}+\acute{\theta}^{2})+V(R)\right]+\left[6A_{1}B_{1}R^2 \dot{\theta}^{2}
{\cal K}_{1}^{2}\cosh(B_{1}{\cal K}_{1}^3)-A_{1}\sinh(B_{1}{\cal K}_{1}^3)\right]+  \nonumber\\&&
\quad\quad \quad \quad\left[6A_{2}B_{2}(\dot{R}^{2}+R^2\dot{\theta}^{2}){\cal K}_{2}^{2}\cosh(B_{2}{\cal K}_{2}^3)-A_{2}\sinh(B_{2}{\cal K}_{2}^3)\right]+ \nonumber\\&&
\quad\quad \quad \quad \left[6A_{3}B_{3}(\dot{R}+R\dot{\theta})^{2}{\cal K}_{3}^{2}\cosh(B_{3}{\cal K}_{3}^3)-A_{3}\sinh(B_{3}{\cal K}_{3}^3)\right]=\varepsilon_{o}+\varepsilon_{1}+\varepsilon_{2}+\varepsilon_{3}.
\end{eqnarray}

An arbitrary hypothetical variation for the non-moving SSWS (\ref{SS})  can be introduced as follows:
\begin{equation}\label{var}
      R(x)=R_{s}+\delta R=0.6\sech(0.6 x)+\xi~ \exp{(-x^{2})},\quad\quad \theta(t)=\omega_{s}t,
\end{equation}
in which $\xi$ is  a small coefficient. Larger  $\xi$ is related to larger variations for the modulus  function.
We consider the phase function to be fixed at $\theta(t)=\omega_{s}t$.  Now, the total energy density (\ref{MTE}) for this  variation (\ref{var})  is reduced    to
\begin{eqnarray}\label{en}
 && \varepsilon(x,t)=\left[\acute{R}^{2}+R^2(\omega_{s}^{2})-R^4+R^3+10R^2\right]+  \left[6A_{2}B_{2}\omega_{s}^{2}R^2{\cal K}_{2}^{2}\cosh(B_{2}{\cal K}_{2}^3)-A_{2}\sinh(B_{2}{\cal K}_{2}^3)\right]+ \nonumber\\&&
 \quad\quad\quad\quad\left[6A_{3}B_{3}\omega_{s}^{2}R^2{\cal K}_{3}^{2}\cosh(B_{3}{\cal K}_{3}^3)-A_{3}\sinh(B_{3}{\cal K}_{3}^3)\right].
\end{eqnarray}
\begin{figure}[ht!]
  \centering
  \includegraphics[width=100mm]{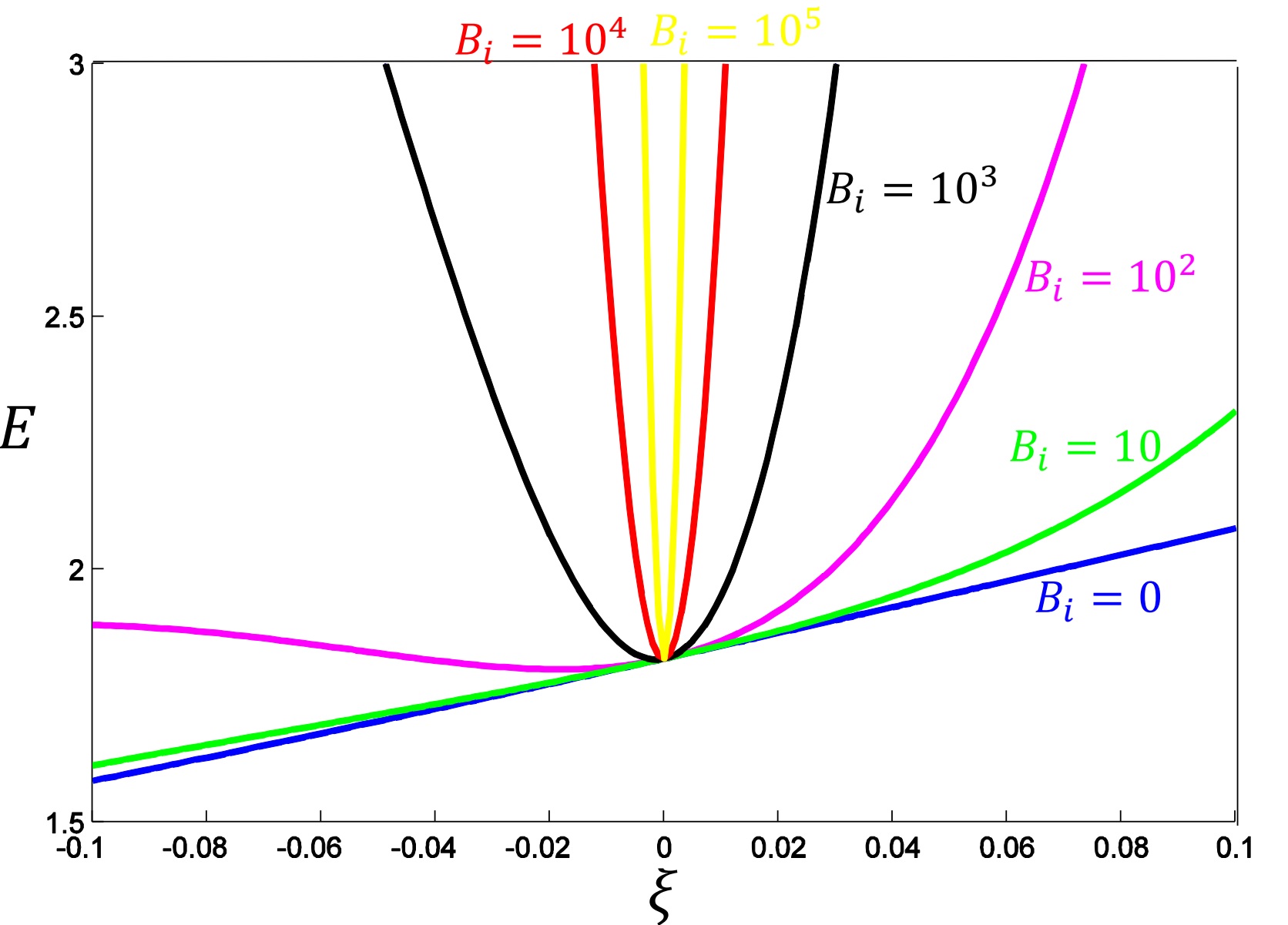}
  \caption{The variations  of  the total rest energy $E_{o}$ versus small $\xi$ for different $B_{i}$'s if one considers an arbitrary  deformation in  the module function of the SSWS (\ref{SS}) according  to Eq.~(\ref{var}). We have set   $A_{i}=1$ ($i=1,2,3$).} \label{no11}
\end{figure}
\begin{figure}[ht!]
  \centering
  \includegraphics[width=100mm]{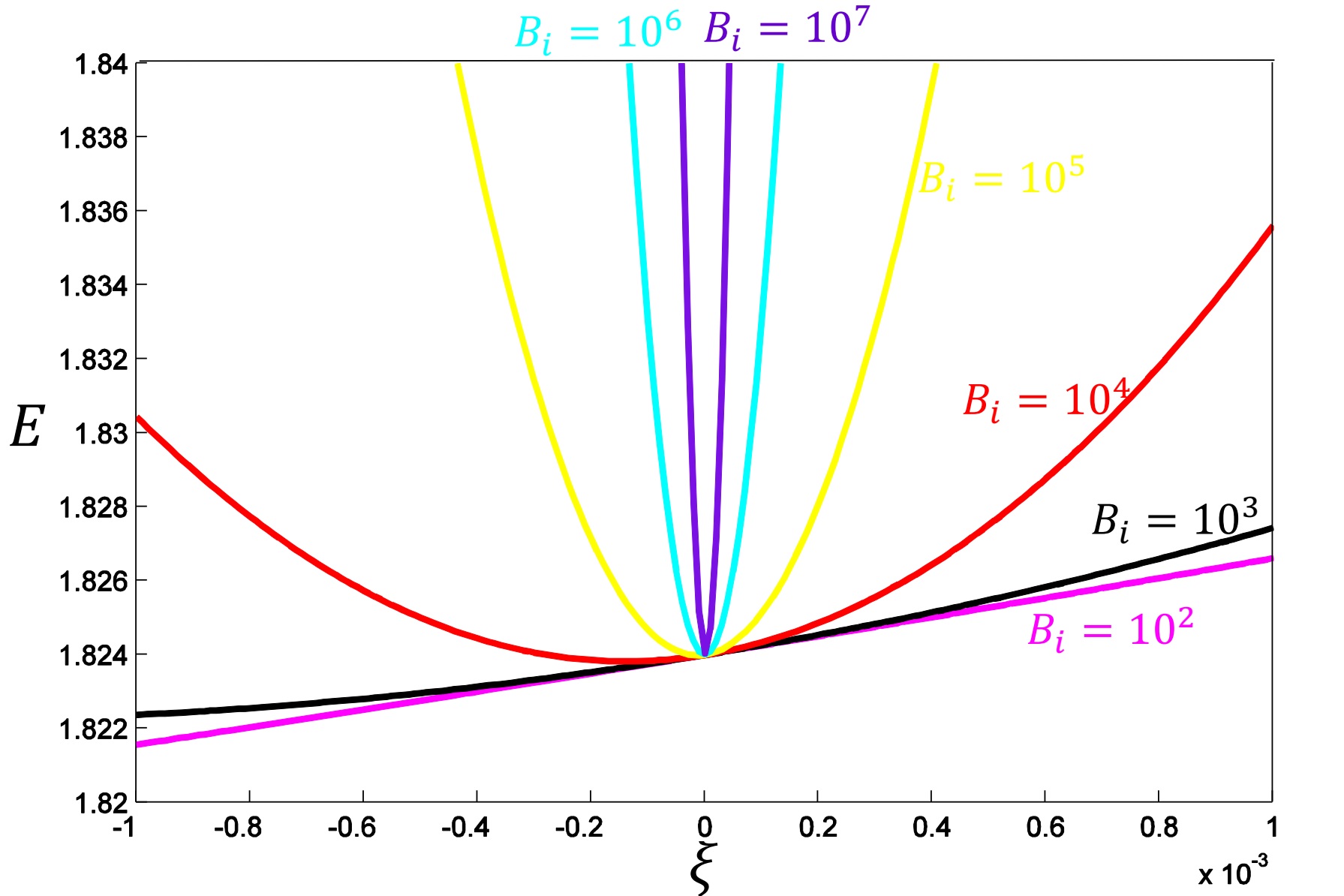}
  \caption{ The variations  of  the total rest energy $E_{o}$ versus small $\xi$ for different $B_{i}$'s  if one considers an arbitrary  deformation in  the module function of the SSWS (\ref{SS}) according  to Eq.~(\ref{var}). We have set  $A_{i}=1$ ($i=1,2,3$). Compare this figure  with   Fig.~\ref{no11}.} \label{no12}
\end{figure}
Note that for this arbitrary variation (\ref{var}): $\dot{R}=0$, $\dot{\theta}^2=\omega_{s}^2=0.64$, $\acute{\theta}=0$ and then ${\cal K}_{1}=0$. The integration of $\varepsilon(x,t)$ (\ref{en}) over all space from $-\infty$ to $+\infty$ yields the total energy ($E$) which is a function of  $\xi$.   The total  energy of the non-deformed SSWS (\ref{SS}) is  $E_{o}=E(\xi=0)=\frac{228}{125}= 1.824$.
As we can see in the  Fig.~\ref{no11}, for small  $B_{i}$'s (i.e. $B_{i}=10$ and $B_{i}=100$), clearly $E(\xi=0)$  is not  a minimum, but by increasing  $B_{i}$'s this behavior fades away slowly, i.e. $E(\xi=0)$,  when we used large $B_{i}$'s (i.e. $B_{i}\geqslant10^{3}$), is apparently a minimum. If we zoom on  around the $\xi=0$ for the cases $B_{i}\geqslant 10^3$, the output result  can be seen  in  the Fig.~\ref{no12}. As we see, for smaller  $\rvert\xi\rvert$, $E(\xi=0)$ is not really a minimum for the cases $B_{i}=10^3$ and $B_{i}=10^4$. Again,  by  increasing $B_{i}$'s, this behavior fades away slowly  and this routine continues in the same way. In other words,   we can always find  a very small range for the coefficients $\xi$ around $\xi=0$, where $E(\xi=0)$ is not  a minimum. This range for larger $B_{i}$'s is apparently smaller. Note that, the same  results can be obtained for the large values of $A_{i}$'s.

\begin{figure}[ht!]
  \centering
  \includegraphics[width=160mm]{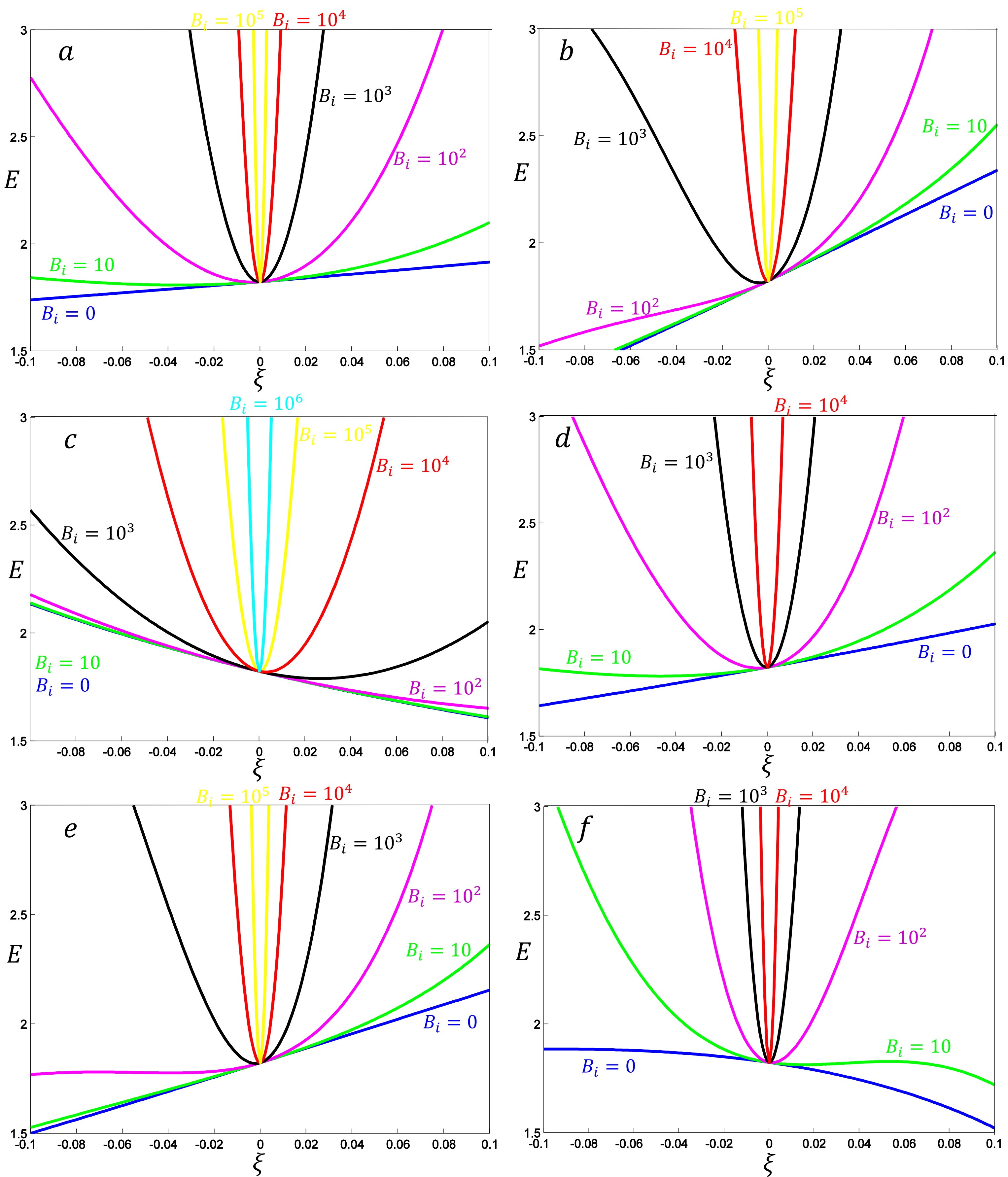}
  \caption{Variations  of  the total rest energy $E_{o}$ versus small $\xi$  and different $B_{i}$'s at $t=0$. We have set   $A_{i}=1$. The Figs a-f are related to different variations (\ref{var1})-(\ref{var6}), respectively.} \label{majmo}
\end{figure}

 Therefore, mathematically,  the SSWS (\ref{SS}) is not really a stable object, but physically, if we consider    large values  for  $B_{i}$'s, there  will be a very small shift from $E(\xi=0)$ which is completely  unimportant and the  stability of the  solitary wave solutions is enhanced appreciably. Therefore, with a very good approximation, we can consider the single solitary wave  solution (\ref{SS}) as a stable object. This treatment is observed for a special hypothetical Gaussian variation for the $R$ function (\ref{var}), though   it is  independent of the form of variations. To support this claim,  we  can study  six other  arbitrary hypothetical variations which are introduced  in the following forms:
 \begin{equation}\label{var1}
R=\omega_{s}'\sech(\omega_{s}' x),\quad  \quad \theta=\omega_{s}t+\xi t ~e^{-x^2},
 \end{equation}
\begin{equation}\label{var2}
R=(\omega_{s}'+\xi)\sech(\omega_{s}' x),\quad  \quad \theta(t)=\omega_{s}t,
\end{equation}
\begin{equation}\label{var3}
R=\omega_{s}'\sech((\omega_{s}'+\xi) x),\quad  \quad \theta(t)=\omega_{s}t,
\end{equation}
\begin{equation}\label{var4}
R=\omega_{s}'\sech(\omega_{s}' x),\quad  \quad \theta=(\omega_{s}+\xi) t,
\end{equation}
\begin{equation}\label{var5}
R=\omega_{s}'\sech(\omega_{s}' x)+\frac{\xi}{1+x^2}\cos(t),\quad  \quad \theta=\omega_{s}t,
\end{equation}
\begin{equation}\label{var6}
R=\sqrt{1-(\omega_{s}+\xi)^2}\sech(\sqrt{1-(\omega_{s}+\xi)^2} x),\quad  \quad \theta=(\omega_{s}+\xi) t.
\end{equation}
All of these variations  for $\xi=0$ turn to the  same non-moving undeformed  SSWS (\ref{SS}).
 The expected results for the variations of the total  energy $E$ versus $\xi$,  for  six   arbitrary deformations (\ref{var1})-(\ref{var6})  at $t=0$, are shown in the Fig.~\ref{majmo} respectively. Note that, the case $B_{i}=0$ is related to the same original standard CNKG system  (\ref{lag}) with the potential (\ref{po}),  and it is quite clear that this case  is by no means stable  according to the new criterion.

In short,  if constants $A_{i}$'s and $B_{i}$'s are considered to be large numbers, the new additional term (\ref{F}) behaves like a zero rest mass spook which surrounds the SSWS (\ref{SS}) and  resists any arbitrary deformation.
In fact, it causes to have  a frozen or rigid  solitary wave  solution (\ref{SS}) for which the  modulus and phase  functions   freeze to $R(x,t)=\omega_{s}'\sech(\omega_{s}'\gamma(x-vt))$ and $\theta(x,t)=k_{\mu}x^{\mu}=\gamma\omega_{s}(t-vx)$, respectively; and the related dominant dynamical equations are the same known standard  versions (5) and (6).


\section{Collisions}\label{sec6}

\begin{figure}[ht!]
  \centering
  \includegraphics[width=160mm]{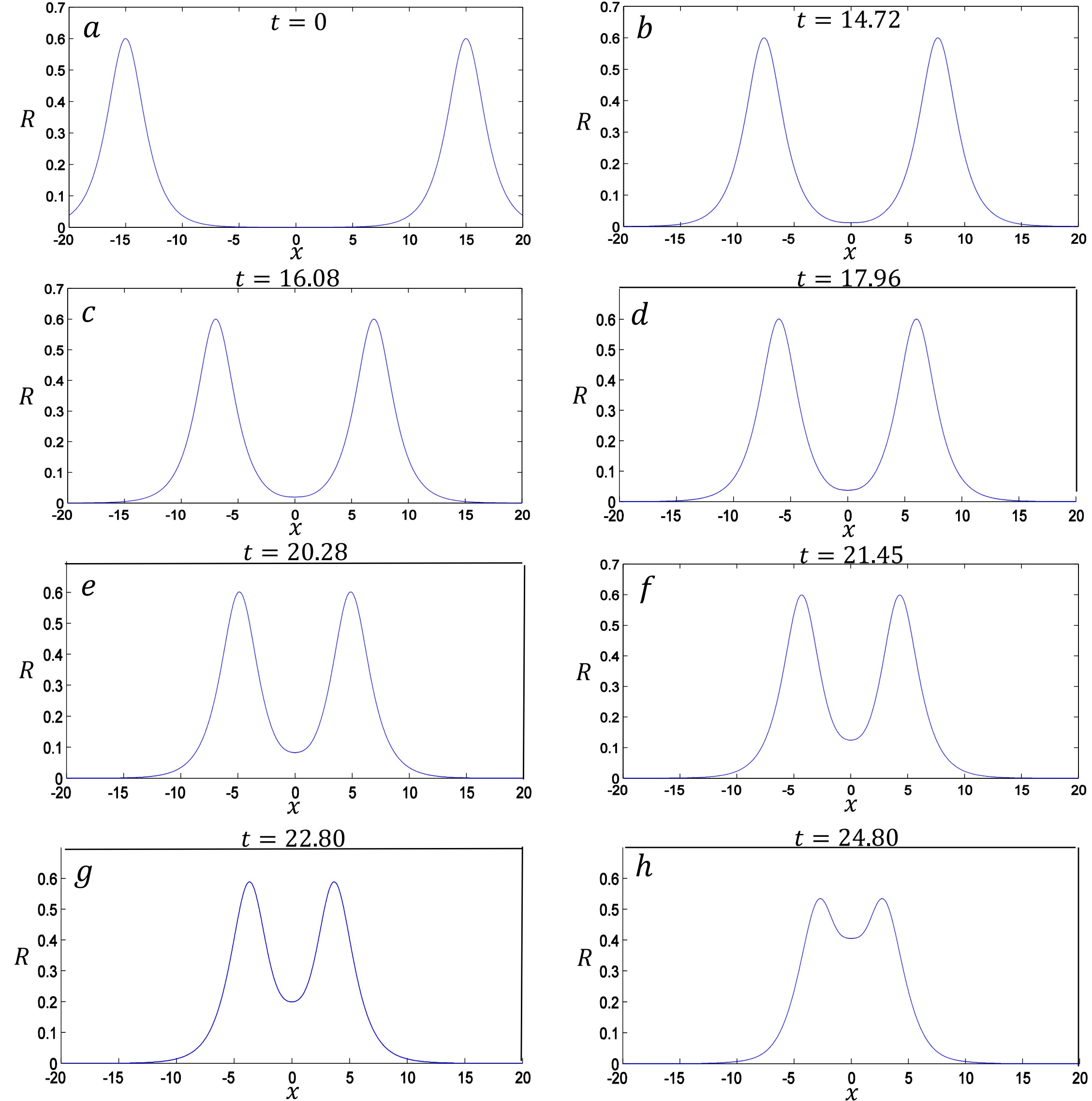}
  \caption{ The module representation of two SSWSs  which are initialized to collide with each other at the same speed  $v=0.5$ for eight different times. We have used the original CNKG system (i.e. $B_{i}=0$)} \label{1}
\end{figure}

In general, a multi lump solution can be constructed easily just by adding single    SSWSs  when they are sufficiently far from each other. In the new extended model ($\ref{lag2}$), the dynamical equations (\ref{geq}) and ($\ref{geq2}$) are  too complicated to be numerically considered. However, based on  the numerical results that obtained  form  the simple  case $B_{i}=0$ ($i=1,2,3$),
we can bring up some statements about the collisions fates in the new extended system (\ref{lag2}) with large  $B_{i}$'s.  For two  SSWSs  which are initialized with the same speed to collide with each other, undoubtedly, their profiles would  change (a little or a lot) when they  approach  each other. We expect the possible changes in the profiles  of the SSWSs,  would be approximately similar to  those  of  the simple case $B_{i}=0$ which are  seen  in the Fig.~\ref{1}.
Now, for different systems with different parameters $B_{i}$'s, if the total energy   is calculated numerically for each profile, it can be shown  that for larger $B_{i}$'s the possibility that two SSWSs   get  closer to each other would be smaller (Table.~\ref{aa}).

The total energy of two far apart   SSWSs, when they move at the same speed of $v=0.5$, is $E=2\gamma E_{o}\thickapprox 4.21$. But,  if they want to get  close to each other with  a finite distance, depending on how large $B_{i}$'s are, they require more initial  energy to occur. Accordingly, if parameters $B_{i}$'s are considered to be  large numbers, we expect two SSWSs to interact with each other through their tail and then reappear after collisions, i.e. essentially they can never be too close together.  In fact, the possible  changes in a SSWS  just occurred for  the energetic collisions, i.e. the collisions for which the speed of the SSWSs are very close to light.

 \begin{table}
 \caption{If the various  profiles which are shown in the Fig.~\ref{1} are considered as the approximations  of the profiles  of two SSWSs for  other systems (\ref{LN}) with different  $B_{i}$'s ($i=1,2,3$), when they  approach   each other, they lead to different  total energies. We have set $A_{i}=1$.}\label{aa}
 \centering
 \begin{center}
 \begin{tabular}{ | l | l | l | l | l | l | l | l | l |}
     \hline
        \multirow{2}{*}{~systems} & \multicolumn{8}{|c|}{profiles} \\ \cline{2-9}
          &\quad ~   a &\quad ~ b &\quad ~ c &\quad ~ d &\quad ~ e &\quad ~ f &\quad ~ g &\quad ~ h \\ \cline{2-9}
        $ B_{i}=10^{8}$ &$E\approx4.21$ & $E\approx10.1$ & $E\approx62.9$ & $E\thickapprox587$&$E\backsim10^5$&    $E\backsim10^{50}$   &$E\backsim\infty$& $E\backsim\infty$ \\
        $ B_{i}=10^{7}$ &$E\approx4.21$ & $E\approx4.8$ &  $E\approx10.1$&$E\approx62.4$&$E\thickapprox604$&$E\backsim10^6$       &$E\backsim10^{106}$     &$E\backsim\infty$ \\
        $B_{i}=10^{6}$ &$E\approx4.21$ & $E\approx 4.3$ &  $E\approx4.8$&$E\approx10.0$&$E\approx63.0$&$E\thickapprox637$&$E\backsim 10^{12}$     &$E\backsim 10^{256}$ \\
        $ B_{i}=10^{5}$ &$E\approx4.21$ & $E\approx4.21$ &  $E\approx4.3$& $E\approx4.8$&$E\approx10.1$&$E\approx61.8$&$E\thickapprox 1500$       &$E\backsim 10^{28}$\\
        $ B_{i}=10^{4}$ &$E\approx4.21$ & $E\approx4.21$ & $E\approx4.21$& $E\approx4.3$&$E\approx4.8$&$E\approx10$&$E\approx63.5$         &$E\sim 10^{4}$ \\
        $ B_{i}=10^{3}$ &$E\approx4.21$ & $E\approx4.21$ & $E\approx4.21$& $E\approx4.21$&$E\approx4.21$&$E\approx4.8$&$E\approx10.1$         &$E\approx66.5$ \\
        $ B_{i}=10^{2}$ &$E\approx4.21$ & $E\approx4.21$ & $E\approx4.21$& $E\approx4.21$&$E\approx4.21$&$E\approx4.3$&$E\approx4.8$         &$E\approx10$ \\
        $ B_{i}=10^{1}$ &$E\approx4.21$ & $E\approx4.21$ & $E\approx4.21$& $E\approx4.21$&$E\approx4.21$&$E=4.21$&$E\approx4.3$         &$E\approx4.8$ \\
        $ B_{i}=0$ &$E\approx4.21$ & $E\approx4.21$ & $E\approx4.21$ &$E\approx4.21$&$E\approx4.21$&$E\approx4.21$&$E\approx4.21$&$E\approx4.21$ \\ \hline
    \end{tabular}
 \end{center}
 \end{table}


In general, for any arbitrary profile, the part of the energy density  that belonged  to the additional term $F$ (i.e. $\sum_{i=1}^{3}\varepsilon_{i}$), would be  always a large positive definite non-zero function,  except for the far apart  SSWSs profiles. In a collision process, while the SSWSs are far away from  each other and then their   profiles are independently unchanged,   the role of the spook term $F$ is zero (i.e. $\sum_{i=1}^{3}\varepsilon_{i}=0$), but when they get  close  to each other and then their profiles change slightly, the role of the spook term becomes  important and  strongly opposes  a closer approach and more changes in the profiles of the SSWSs. For example, according to Fig.~{1} and Table.~\ref{aa}, if we consider a system with $B_{i}=10^{8}$, to put two SSWSs at an approximate distance of 10, the initial energy  must be in the order of $10^{5}$ or the initial speed must be approximately equal to $0.999999999$. Therefore, we can be sure that for the systems with large $B_{i}$'s, there is always a huge repulsive force between SSWSs which not allow two distinct SSWSs  to get close  together. Hence, we  expect  they reappear with no considerable changes after collisions.

 If we consider the systems  for which  parameters  $B_{i}$'s (or $A_{i}$'s)  be extremely large numbers,  we can divide the nature of such systems into two distinct stationary  parts: first, the vacuum state, and second, the free far apart SSWSs. Except the free far apart SSWSs and the vacuum state ($R=0$), for  other possible stable  field  solutions (structures), always $\sum_{i=1}^{3}\varepsilon_{i}$ would be a very large  positive definite function which yields a very large total energy, then infinite energy is required for them to be created.

\section{Summary and conclusion}\label{sec7}

We first reviewed some basic properties of the complex nonlinear KG (CNKG) systems in $1+1$ dimensions. Each CNKG system may have some  non-dispersive solitary wave  solutions with  particular  rest frequencies  ($\omega_{o}$) and rest energies ($E_{o}$),  called Q-balls. Traditionally, two distinct criteria    are used to check the stability of the Q-balls: the classical  criterion and the quantum mechanical criterion. In this paper, we used  a new criterion for examining  the stability (i.e. the energetically stability criterion) of a solitary wave solution that is based on  examining   the changes in the total energy  for  arbitrary   small variations   above  the background of  the  special solitary wave solution. In other words,  a   special solitary wave solution  is energetically stable, if the total energy, for any arbitrary variation in its internal structure, always increases. Accordingly, we showed that  in general, there is not any CNKG system with an energetically  stable Q-ball solution at all.



Inspire by the well-known quantum field theory  in which any standard  Lagrangian density is (nonlinear) Klein-Gordon (-like) and is used just for a special type of known particles with specific properties, classically we assume that there is a new  extended CNKG system with a single stable solitary wave  solution (Q-ball) for which the general dynamical equations (and the other properties) are reduced to those versions of a  standard CNKG system.
 In fact, we put forward three  basis postulates. First, we assumed a relativistic localized wave  function (\ref{SS}) as a single hypothetical particle of an unknown field model. Second, we assumed that   the dominant dynamical  equations of motion  just for this special  solution (\ref{SS}) are the same  standard known CNKG versions. And eventually we assumed  that this special solution (\ref{SS}) is an energetically  stable  solution.
All of these  postulates  oblige  us  to add a proper  term $F$ to the original CNKG Lagrangian density, where it and all of its derivatives should   be zero for this special solitary wave    solution (SSWS) (\ref{SS}).



In this regard, it was introduced three independent functional scalars $\mathbb{S}_{i}$ ($i=1,2,3$), which  are zero simultaneously just for the trivial vacuum state $R=0$ and a non-trivial SSWS (\ref{SS}).  In other words, the SSWS (\ref{SS}) is the unique non-trivial  common  solution of three independent conditions $\mathbb{S}_{i}$'s$=0$. The proper additional term $F$,  which is considered  in the new extended CNKG model (\ref{lag2}),
 can be  considered in the following form:  $F=\sum_{i=1}^{3} A_{i}f(Z_{i})$, where $Z_{i}=B_{i}{\cal K}_{i}^n$, $n=3,5,7,\cdots$,  $A_{i}$'s and $B_{i}$'s are some positive constants,  $f$ is any arbitrary  odd $\sinh$-like function, and ${\cal K}_{i}$'s are  three special independent linear combinations of  $\mathbb{S}_{i}$'s.
For such  proper additional terms (\ref{F}), the corresponding   energy density function (\ref{MTEi}) is decomposed into four distinct parts $\varepsilon_{i}$ ($i=0,1,2,3$). In general, $\varepsilon_{i}$'s ($i=1,2,3$)  are positive definite functions, that any of their terms contains one of the even  powers of  ${\cal K}_{i}$'s, and are zero simultaneously just for the non-trivial SSWS (\ref{SS}) and trivial vacuum state $R=0$. Except $\varepsilon_{o}$, which originates from the basic standard CNKG system (\ref{lag}), the other parts of the energy density function, i.e. $\varepsilon_{i}$ ($i=1,2,3$), all originate from the additional term $F$ and all  contain parameters   $B_{i}$'s and $A_{i}$'s ($i=1,2,3$).   If parameters $B_{i}$'s and $A_{i}$'s are considered  to be large numbers, thus   $\varepsilon_{i}$'s ($i=1,2,3$) are large functions in compared with function $\varepsilon_{o}$. More precisely, for the other solutions of the system, which are not very close to the trivial vacuum state $R=0$ and non-trivial SSWS (\ref{SS}), always  at least one of the independent functionals  ${\cal K}_{i}$'s  is not   zero, and then  at least one of the $\varepsilon_{i}$ ($i=1,2,3$) is a large non-zero positive function. Accordingly,   it was shown analytically and numerically that   the SSWS (\ref{SS}) would be approximately  an energetically stable solution, provided   $B_{i}$'s or $A_{i}$'s ($i=1,2,3$) are considered  to be  large number. In fact, there are always very small arbitrary variations above the background of the SSWS (\ref{SS}) for which the total energy decreases. But, this decreasing is so small that can be physically ignored in the stability considerations. However, for the other significant  small variations, it was shown that  the total energy always increases and the energetically stability of the SSWS (\ref{SS}) would guaranteed appreciably.

The stability for the   SSWS (\ref{SS}) would be intensified by taken into account the larger values of parameters   $B_{i}$'s or $A_{i}$'s ($i=1,2,3$) which appeared in the new additional term $F$. In other words, the larger the values the greater will be the increase in the total energy for any arbitrary small variation above the background of SSWS (\ref{SS}). Accordingly, the proper additional term $F$ (\ref{F}) behaves like a massless spook which surrounds the single SSWS (\ref{SS}) and  resists any arbitrary significant small deformations in  its internal structure. The role of the additional term $F$ in the collisions behaves  like a huge repulsive force which does  not allow  two SSWSs to get close each other. Therefore, it is expected that SSWSs reappear without any distortion in collisions with   each other.



If one considers  a system  for which  parameters  $B_{i}$'s (or $A_{i}$'s)  be extremely large numbers, then the other configurations of the fields $R$ and $\theta$, which are not very close to any number of distinct  far apart SSWSs and trivial vacuum state $R=0$, require  infinite external energy  to be created. In other words, if one considers this system as a real physical
system, since it is not possible to provide an extremely large external energy at a special
place for creating the other  configurations of the fields $R$ and $\theta$, thus the only non-trivial configurations of the fields with the finite energies would be any number of the far apart SSWSs  as a multi particle-like solution. Physically this issue can be
interesting, in fact it classically explains how a system leads to many identical particles with the specific characteristics. In fact, the free far apart SSWSs can be called the quanta of the system classically.


To summarize, this paper introduces  an extended CNKG system that  yields a single non-topological energetically stable solitary wave solution for which the general dynamical equations are  reduced to  those versions of a special type of the standard well-known CNKG systems. It is noteworthy to mention, only  some relativistic topological solutions such as kinks (antikinks) have been introduced as  energetically stable objects so far, but the existence of a relativistic energetically stable  non-topological solution has not been previously reported (at least as far as we searched) and this work introduces a new one (\ref{SS}). Moreover,  for other forthcoming works, especially in $3+1$ dimensions, it has been attempted to accurately provide  all the mathematical tools required  in this paper. For example, we hope to write a series of articles in near future that classically explains how the universal constant $\hbar$ can be justified for all particles, and the mathematical tools presented in this article are very important for achieving this goal.


\appendix

\section{}

Here, we are going to show that three PDEs
\begin{eqnarray} \label{true1}
  &&\mathbb{S}_{1}=\dot{\theta}^2-\theta'^2-\omega_{s}^2=0,\\\label{true2}&&
 \mathbb{S}_{2}=\dot{R}^2-R'^2+V(R)-\omega_{s}^2 R^2=0,\\\label{true3}&&
 \mathbb{S}_{3}=\dot{R}\dot{\theta}-R'\theta'=0.
\end{eqnarray}
do not have any non-trivial common solution except the SSWS (\ref{SS}). Equation (\ref{true3}) leads to obtain $\dot{\theta}$ in terms of $\theta'$, $R'$ and $\dot{R}$ as follows:
\begin{equation}\label{fgh}
\dot{\theta}=\dfrac{R'\theta'}{\dot{R}}.
\end{equation}
If we insert this into Eq.~(\ref{true1}), we can obtain $\theta'$ in terms of $\varphi'$ and $\dot{\varphi}$ as follows:
\begin{equation}\label{fgh2}
\theta'=\dfrac{\omega_{s}\dot{R}}{\sqrt{R'^2-\dot{R}^2}},
\end{equation}
where $\omega_{s}=\pm 0.8$. Using Eqs. (\ref{fgh}) and (\ref{fgh2}), $\dot{\theta}$ can be obtained  as well:
\begin{equation}\label{fgh3}
\dot{\theta}=\dfrac{\omega_{s} R'}{\sqrt{R'^2-\dot{R}^2}}.
\end{equation}
The obvious mathematical  expectation $(\dot{\theta})'=\dfrac{d}{dx}\dfrac{d\theta}{dt}=\dfrac{d}{dt}\dfrac{d\theta}{dx}=\dot{(\theta')}$
leads to the following result:
\begin{equation}\label{vgh}
\ddot{R}-R''+\dfrac{1}{\sqrt{R'^2-\dot{R}^2}}
(\dot{R}^2\ddot{R}+R'^2 R''-2\dot{R}R'\dot{R}')=0,
\end{equation}
which simply can be written  in a covariant form:
\begin{equation}\label{fjh}
\partial_{\mu}\partial^{\mu}R+\frac{1}{\sqrt{-\partial_{\mu}R\partial^{\mu}R}}(\partial_{\nu}R\partial_{\sigma}R)
(\partial^{\nu}\partial^{\sigma}R)=0
\end{equation}
Therefore, to find  the common solutions of three independent nonlinear PDEs (\ref{true1}), (\ref{true2}) and (\ref{true3}), equivalently,   we can  search  for the common solutions of the    two different PDEs (\ref{true2}) and (\ref{fjh}).  In general, it is easy to show that each non-vibrational   function  $R_{v}(x,t)=R_{o}(\gamma(x-vt))$, would be a solution of the  PDE (\ref{fjh}) or (\ref{vgh}). Moreover, for any non-vibrational solitary wave solution, Eqs. (\ref{fgh2}) and (\ref{fgh3}) lead to $\theta'=\omega_{s}\gamma v=\omega v$ and $\dot{\theta} =\gamma\omega_{s}=\omega$ as we expected.
On the other hand, we know that the SSWS (\ref{SS}) is the single  non-vibrational localized  solution of the PDE (\ref{true2}). Hence, for PDEs (\ref{true2}) and (\ref{fjh}), the single common non-vibrational  localized solitary wave solution is the same  SSWS (\ref{SS}), as we expected. Accordingly,  for the module field $R$, there are two completely different PDEs (\ref{true2}) and (\ref{fjh}). Hence it does seem that there are  other common vibrational localized solutions along with the non-vibrational SSWS (\ref{SS}).


\begin{thebibliography}{99}
\bibitem{rajarama} R. Rajaraman, \textit{Solitons and Instantons }(North Holland, Elsevier, Amsterdam, 1982).
\bibitem{Das}  A. Das, \textit{Integrable Models} (World Scientific, 1989).
\bibitem{lamb} G. L. Lamb, \textit{Elements of Soliton Theory} (Dover Publications, 1995).
\bibitem{Drazin} P. G. Drazin and R. S. Johnson, \textit{Solitons: an Introduction} (Cambridge University Press,
1989).
\bibitem{TS} N. Manton,  P. sutcliffe,  \textit{Topological Solitons}, (Cambridge University Press, 2004).


\bibitem{phi41} D. K. Campbell and M. Peyrard,  Physica D, \textbf{19}, 165 (1986).
\bibitem{phi42} D. K. Campbell and M. Peyrard,  Physica D, \textbf{18}, 47 (1986).
\bibitem{phi43} D. K. Campbell, J. S. Schonfeld, and C. A. Wingate,  Physica D, \textbf{9}, 1 (1983).
\bibitem{phi44} M. Peyrard and D. K. Campbell, Physica D, \textbf{9}, 33 (1983).
\bibitem{phi45}R. H. Goodman and R. Haberman, Siam J. Appl. Dyn. Syst., \textbf{4}, 1195 (2005).
\bibitem{OV} O. V. Charkina, M. M. Bogdan, Symmetry Integr.  Geom., \textbf{2}, 047 (2006).
    \bibitem{GH} A. R. Gharaati, N. Riazi and F. Mohebbi, Int. J. Theor. Phys., \textbf{45}, 53 (2006).
\bibitem{MM1}  M. Mohammadi and N. Riazi, Prog. Theor. Phys., \textbf{126}, 237 (2011).
\bibitem{MR}    M. Mohammadi and N. Riazi, Commun. Nonlinear Sci. Numer. Simul., \textbf{72}, 176-193 (2019).
\bibitem{JRM1} J. R. Morris, Annals of Physics, \textbf{393},  122-131 (2018).
\bibitem{JRM2} J. R. Morris, Annals of Physics, \textbf{ 400}, 346-365 (2019).





\bibitem{DSG1} V. A. Gani and A. E. Kudryavtsev, Phys. Rev. E, \textbf{60}, 3305 (1999).
\bibitem{DSG2} C. A. Popov, Wave Motion, \textbf{42}, 309 (2006).

\bibitem{DSG3} M. Peyravi, A. Montakhab, et al., Eur. Phys. J. B, \textbf{72}, 269 (2009).

\bibitem{MM2} M. Mohammadi, N. Riazi, and A. Azizi, Prog. Theor. Phys., \textbf{128}, 615 (2012).
\bibitem{waz} A. M. Wazwaz, Chaos, Solitons and Fractals, \textbf{28}, 1005 (2006).
\bibitem{ana1} H. Hassanabadi, L. Lu,  et al., Annals of Physics,
\textbf{342}, 264-269 (2014).
\bibitem{ana2} A. Alonso-Izquierdoa, J. Mateos Guilarteb, Annals of Physics,
 \textbf{327}, 2251-2274 (2012).





\bibitem{Kink1} P. Dorey, K. Mersh, T. Romanczukiewicz, and Y. Shnir, Phys. Rev. Lett., \textbf{107}, 091602  (2011).
\bibitem{Kink2} V. A. Gani, A. E. Kudryavtsev, and M. A. Lizunova, Phys. Rev. D, \textbf{89}, 125009 (2014).
\bibitem{Kink3} A. Khare, I. C. Christov, and A. Saxena, Phys. Rev. E, \textbf{90}, 023208 (2014).

\bibitem{Kink4} A. Moradi Marjaneh, V. A. Gani, D. Saadatmand, J. High Energ. Phys., \textbf{07}, 028 (2017).

\bibitem{Kink5} D. Bazeia, E. Belendryasova, and V. A. Gani, Eur. Phys. J. C, \textbf{78}, 340 (2018).
\bibitem{Kink6} V. A. Gani, A. Moradi Marjaneh, et al., Eur. Phys. J. C, \textbf{78}, 345 (2018).
\bibitem{Kink7} P. Dorey and T. Romańczukiewicz, Physics Letters B, \textbf{779}, 117-123 (2018).
\bibitem{Kink8} I. C. Christov, R. J. Decker,  et al., Phys. Rev. D, \textbf{99}, 016010 (2019).
\bibitem{Kink9} E. Belendryasova and V. A. Gani, Commun. Nonlinear Sci. Numer. Simul., \textbf{67}, 414-426 (2019).

\bibitem{Kink10} D. Bazeia, R. Menezes, and D. C. Moreira, J. Phys. Commun., \textbf{2}, 055019 (2018).

\bibitem{Kink11}  I. C. Christov, R. J. Decker,  et al., Phys. Rev. Lett., \textbf{122}, 171601 (2019).
\bibitem{Kink12}  N. S. Manton, J. Phys. A: Math. Theor., \textbf{52}, 065401 (2019).


\bibitem{Kink13}   V. A. Gani, V. Lensky, and M. A. Lizunova,  J. High Energ. Phys., \textbf{08}, 147 (2015).

\bibitem{Kink14}  V. A. Gani, M. A. Lizunova, and R. V. Radomskiy, J. High Energ. Phys., \textbf{04}, 043 (2016).



\bibitem{Kink15}   D. Bazeia, A. R. Gomes, et al., Physics Letters B, \textbf{793},  26-32 (2019).
 \bibitem{Kink16}  V. A. Gani, A. Moradi Marjaneh, and D. Saadatmand, Eur. Phys. J. C,  \textbf{79}, 620 (2019).
\bibitem{Kink17} D. Bazeia, A. R. Gomes, et al.,  Int. J. Mod. Phys. A, \textbf{34}, 1950200 (2019).


\bibitem{toft} G. 't Hooft, Nuclear Physics B, \textbf{79},  276 (1974).
\bibitem{MKP} M. K. Prasad, Physica D,  \textbf{1}, 167-191 (1980).
\bibitem{TOF} S. Nishino, R. Matsudo, et al.,  Prog. Theor. Exp. Phys.,  \textbf{2018}, 103B04 (2018).


\bibitem{SKrme} T. H. R. Skyrme,  Proc. Roy. Soc. A, \textbf{260}, 127 (1961).
\bibitem{SKrme2} T. H. R. Skyrme,  Nucl. Phys., \textbf{31}, 556 (1962).
\bibitem{SKrme3} N. S. Manton, B. J. Schroers, and M. A. Singer, Commun. Math. Phys., \textbf{245}, 123-147 (2004).
\bibitem{SKrme4} N. S. Manton, Commun. Math. Phys., \textbf{111}, 469 (1987).


\bibitem{Vak1} N. G. Vakhitov and A. A. Kolokolov, Radiophys Quantum Electron, \textbf{16}, 783 (1973).
\bibitem{Vak2} A. A. Kolokolov, J. Appl. Mech. Tech. Phys., \textbf{14}, 426 (1973).
\bibitem{Vak777} R. Friedberg, T. D. Lee and A. Sirlin, Phys. Rev. D, \textbf{13}, 2739 (1976).
\bibitem{Vak3} A. G. Panin, and M. N. Smolyakov, Phys. Rev. D, \textbf{95}, 065006 (2017).
\bibitem{Vak4} A. Kovtun, E. Nugaev, and A. Shkerin, Phys. Rev. D, \textbf{98}, 096016 (2018).
\bibitem{Vak5} M. N. Smolyakov, Phys. Rev. D, \textbf{97}, 045011 (2018).
\bibitem{Vak6} M. I. Tsumagari, E. J. Copeland, and P. M. Saffin, Phys. Rev. D, \textbf{78}, 065021 (2008).









\bibitem{Lee3} T. D.  Lee and Y. Pang, Phys. Rep., \textbf{221}, 251 (1992). 
\bibitem{Scoleman} S. Coleman, Nucl. Phys. B, \textbf{262}, 263 (1985). 
\bibitem{R1} D. Bazeia, M. A. Marques, and R. Menezes, Eur. Phys. J. C, \textbf{76}, 241 (2016).
\bibitem{R2} D. Bazeia, L. Losano, et al.,   Physics Letters B, \textbf{765}, 359 (2017). 
\bibitem{R3} K. N. Anagnostopoulos, M. Axenides, et al.,  Phys. Rev. D, \textbf{64}, 125006 (2001).
\bibitem{R4} M. Axenides, S. Komineas, et al., Phys. Rev. D, \textbf{61}, 085006 (2000).
\bibitem{R5} P. Bowcock, D. Foster, and P. Sutcliffe, J. Phys. A: Math. Theor., \textbf{42}, 085403 (2009).
\bibitem{R6} T. Shiromizu, T. Uesugi, and M. Aoki, Phys. Rev. D, \textbf{59}, 125010 (1999).
\bibitem{R7} T. Shiromizu, Phys. Rev. D, \textbf{58}, 107301 (1998).

\bibitem{Riazi2} N. Riazi, Int. J. Theor. Phys., \textbf{50}, 3451 (2011).
\bibitem{MM3} M. Mohammadi and N. Riazi,  Prog. Theor. Exp. Phys., \textbf{2014}, 023A03 (2014).
\bibitem{Derrick} G. H. Derrick, Journal of Mathematical Physics, \textbf{5}, 1252 (1964).

\bibitem{PH1} M. Mohammadi, Iran. J. Sci. Technol. Trans. Sci., \textbf{43}, 2627-2634 (2019).
\bibitem{PH2}  M. Mohammadi and R. Gheisari,
Physica Scripta, \textbf{95}, 015301  (2020).

\bibitem{PH3} J. Diaz-Alonso and D. Rubiera-Garcia, Annals of Physics,  \textbf{324}, 827-873 (2009).






\end{thebibliography}
\end{document}